\documentclass{aa}  

\usepackage{footnote}
\usepackage{verbatim}
\usepackage{booktabs}
\usepackage{graphicx}
\usepackage{subfloat}
\usepackage{siunitx}
\usepackage{afterpage}
\usepackage{textgreek}
\usepackage{siunitx}
\usepackage{float}
\usepackage{lscape}
\usepackage{color}
\usepackage{combelow}
\usepackage{txfonts}
\usepackage{hyperref}

\DeclareSIUnit{\jansky}{Jy}
\DeclareSIUnit{\year}{yr}
\DeclareSIUnit{\parsec}{pc}
\newcommand{\sources}{3C\,196, 3C\,225, 3C\,273, 3C\,295, 3C\,298, and 3C\,380}
\newcommand{\mjybeam}{$\text{mJy beam}^{-1}$}

\begin{document}

   \title{Pushing subarcsecond resolution imaging down to 30 MHz with the trans-European International LOFAR Telescope}
    
   \titlerunning{Subarcsecond resolution ILT images down to 30~MHz}

   \author{C. Groeneveld \inst{1} \thanks{E-mail:\url{groeneveld@strw.leidenuniv.nl}}
          \and
          R.~J. van Weeren \inst{1}
          \and
          G.~K. Miley \inst{1}
          \and
          L.~K. Morabito \inst{2,3}
          \and
          F. de Gasperin \inst{4,5}
          \and
          J.~R. Callingham \inst{1,6}
          \and
          F. Sweijen \inst{1}
          \and
          M. Br\"uggen  \inst{4}
          \and
          A. Botteon \inst{1}
          \and
          A. Offringa \inst{6,7}
          \and
          G. Brunetti \inst{5}
          \and
          J. Moldon \inst{8}
          \and
          M. Bondi \inst{5}
          \and
          A. Kappes \inst{9}
          \and
          H.~J.~A. R\"ottgering \inst{1}
          }

   \institute{\inst{1} Leiden Observatory, Niels Bohrweg 2, 2333 CA Leiden\\
              \inst{2} Centre for Extragalactic Astronomy, Department of Physics, Durham University, Durham DH1 3LE, UK\\
              \inst{3} Institute for Computational Cosmology, Department of Physics, University of Durham, South Road, Durham DH1 3LE, UK\\
              \inst{4} Hamburger Sternwarte, University of Hamburg, Gojenbergsweg 112, D-21029, Hamburg, Germany\\
              \inst{5} INAF - Istituto di Radioastronomia, via P. Gobetti 101, 40129, Bologna, Italy\\
              \inst{6} ASTRON, Netherlands Institute for Radio Astronomy, Oude Hoogeveensedijk 4, Dwingeloo, 7991\,PD, The Netherlands \\
              \inst{7} Kapteyn Astronomical Institute, University of Groningen, PO Box 800, 9700 AV Groningen, The Netherlands \\
              \inst{8} Instituto de Astrof\'isica de Andaluc\'ia (IAA, CSIC), Glorieta de las Astronom\'ia, s/n, E-18008 Granada, Spain \\
              \inst{9} Institut f\"{u}r Theoretische Physik und Astrophysik, Universit\"{a}t W\"{u}rzburg, Emil-Fischer-Str. 31, 97074 W\"{u}rzburg, Germany
             }

   \date{Received XXXX; accepted YYYY}

  \abstract
  {Relatively little information is available about the Universe at ultra-low radio frequencies (ULF; i.e.,   below 50~MHz), although the ULF spectral window contains a wealth of unique diagnostics for studying galactic and extragalactic phenomena.
   Subarcsecond resolution imaging at these frequencies is extremely difficult, due to the long baselines (>1000~km) required and large ionospheric perturbations.
  We have conducted a pilot project to investigate the ULF performance and potential of the International LOFAR Telescope (ILT), a trans-European interferometric array with baselines up to $\sim$2000~km and observing frequencies down to 10~MHz. We have successfully produced images with subarcsecond resolution for six radio sources at frequencies down to 30~MHz. This resolution is  more than an order of magnitude better  than pre-ILT observations at similar frequencies. The six targets that we  imaged (\sources) are bright radio sources with compact structures. By comparing our data of 3C\,196 and 3C\,273 with observations at higher frequencies, we investigate their spatially resolved radio spectral properties. 
  Our success shows that at frequencies down to 30~MHz, subarcsecond imaging with the ILT is possible.
  Further analysis is needed to determine the feasibility of observations of fainter sources or sources with less compact emission.
  }    
    

   \maketitle

\section{Introduction}

Radio observations at decameter wavelengths (ultra-low frequencies, ULF\footnote{Our usage of ULF deviates from the ITU convention. The frequency range used in this paper falls in the VHF (very high-frequency) band, but we have chosen to use ULF to differentiate from regular low frequencies observed with LOFAR}, <50~MHz) were among the first observations done in the history of radio astronomy.
However,  at decameter wavelengths the sky has  not been explored to the same depths as at gigahertz frequencies.
Interferometric observations at ULF are challenging for several reasons.
First and most important, the ionosphere plays a major role in corrupting astronomical signals at ULF \citep[e.g.,][]{intema_ionosphere,1986isra.book.....T,sleewraith}.
Second, longer physical baselines are required to achieve arcsecond resolution at ULF compared to higher frequencies.

\citet{sleewraith} showed that fringes could be obtained at 38~MHz over a 120~km baseline, indicating that high-resolution  ULF observations are, in principle, feasible.
More recently, the Ukrainian Radio Interferometer (URAN) has found fringes over baselines of $\sim$900~km at 25~MHz \citep{uran_169,uran_380}.
However, neither of these interferometers have sufficient $uv$-coverage and sensitivity for ULF imaging to be attempted.
The International LOFAR Telescope (ILT) has baselines as long as $\sim$ 2000~km \citep{lofar_paper}, which results in an unprecedented spatial resolution of $\sim$ 0.6 arcseconds at a frequency of $50$~MHz.
This is comparable to typical resolutions achieved by GHz frequency instruments, and allows  spatially resolved spectral index analysis.

Achieving high resolutions at low frequencies is important and could be transformational because of the unique diagnostic tools that the ULF provides for studying emission and absorption processes in galactic and extragalactic objects.
One reason is that absorption processes (such as synchrotron self-absorption, a cutoff in low-energy electrons, and free-free absorption) in radio galaxies \citep[e.g., Cygnus A,][]{2016MNRAS.463.3143M} or star-forming galaxies \citep[][]{varenius_m82} are more readily traced at low frequencies \citep{callingham_2017}.
Low-frequency observations provide a large lever arm in frequency space to sample the optically thick part of the spectrum, which is key to differentiating between competing absorption mechanisms \citep{callingham_2015}.
Another reason is that synchrotron aging, which causes steep spectra at high frequencies, means that some sources or components of sources may be observable only at low frequencies.
Examples of such sources include fossil plasma in clusters and aged electrons from old AGN outbursts \citep{Harwood13_ageing,Harwood15_ageing}.
In addition, observing at low frequencies allows access to lower rest frequencies for redshifted objects, which is critical for detecting synchrotron aging or absorption processes in objects at higher redshifts.
Another class of objects of potential interest are those with an intrinsic cutoff  in their spectra.
For example, the Jupiter--Io system can only be seen below 40 MHz \citep{zarka_jupio}, and it has been predicted that some stars and exoplanets are only visible at low frequencies \citep{vedantham_reddwarf,callingham_binary}.
Finally, at ULF, radiation mechanisms such as plasma oscillations produce coherent radiation.
Therefore, ULF observations with the ILT could potentially reveal and study hitherto undetected classes of objects not visible at higher frequencies.\\

The ILT is particularly suited for high-resolution imaging at low frequencies because of the large number of long baselines providing substantial baseline coverage at low frequencies.
Previously, only two targets have been successfully imaged with the large  $uv$-coverage of the ILT at frequencies below 100 MHz.
The target 3C\,196 was imaged over the frequency range of 30--80~MHz \citep{2010iska.meetE..58W} with an early incomplete ILT array, and 4C\,43.15 was imaged between frequencies of 47 and 63~MHz \citep{2016MNRAS.461.2676M}.
Since then, the ILT has been expanded with additional European stations, resulting in further improvements in sensitivity.
Furthermore, the increased number of international stations and recent advances in software for imaging, calibration, and radio frequency interference (RFI) excision with the ILT can help overcome some of the challenges of high-resolution ULF imaging.

In this paper we describe a pilot project to perform high-resolution ULF imaging of six bright compact radio sources (\sources) with the Low Band Antenna (LBA) of the International LOFAR Telescope \citep{lofar_paper}.
We compared images of 3C\,196 and 3C\,273 with Very Large Array (VLA) observations at higher frequencies, where available, in order to construct spectral index maps with arcsecond resolution.
The outline of this paper is as follows. In Sect.~\ref{sec:obsdata} we describe the observations and data reduction. The results are presented in Sect.~\ref{sec:results}. This is followed by a discussion of the future prospects and conclusions in Sects.~\ref{sec:discussion} and~\ref{sec:conclusions}.
Throughout the paper we adopt a \textLambda CDM cosmology \citep{Planck_cosmology}, with $H_0 = \SI{67.4}{\kilo\meter\per\second\per\mega\parsec}$ and $\Omega_M = 0.315$.
We define our spectral index $\alpha$ as $S\sim \nu^\alpha$.

\begin{table*}[h]
    \centering
    \caption{Sources observed, sorted by right ascension. Coordinates and 74 MHz flux densities were obtained from the VLA Low-Frequency Sky Survey Redux \citep{2014MNRAS.440..327L}. The galaxy host types are from \cite{Laing1983} (1), where Q refers to a spectroscopically confirmed quasar and R to a non-quasar galaxy; \cite{1963Natur.197.1040S} (2) and \cite{298quasar} (3).
    The largest angular sizes are   from \cite{Reid_jodrell_bank} at 966~MHz (4), \cite{Jenkins_cambridge} at 5~GHz (5), \cite{2017A&A...601A..35P} at 1.37~GHz (6), \cite{akujor_1994} at 1.6~GHz (7), and \cite{akujor_1995} at 1.6~GHz (8).}
    \begin{tabular}{lrrrrr}
        Source name & RA (J2000) & DEC (J2000) & $S_{74}$ [Jy] & Host type & Largest angular size [arcsec]\\
        \hline
        3C\,196 & 08:13:36.22 & +48:13:02.5 & 130.8 & Q${}^1$ & 10${}^{(4)}$\\
        3C\,225 & 09:42:15.30 & +13:45:49.3 & 32.5  & R${}^1$ & 6${}^{(5)}$ \\
        3C\,273 & 12:29:05.89 & +02:02:55.7 & 161.7 & Q${}^2$ &  22${}^{(6)}$ \\
        3C\,295 & 14:11:20.24 & +52:12:06.6 & 128.9 & R${}^1$ & 4.7${}^{(7)}$ \\
        3C\,298 & 14:19:08.30 & +06:28:33.9 & 95.27 & Q${}^3$ & 1.5${}^{(8)}$\\
        3C\,380 & 18:29:32.49 & +48:44:57.5 & 124.7 & Q${}^1$ & 20${}^{(4)}$
    \end{tabular}
    \label{tab:sources}
\end{table*}

\section{Observations and data reduction}
\label{sec:obsdata}

\subsection{LOFAR data processing}
The six targets observed (\sources; see \autoref{tab:sources}) were selected because of their bright, compact emission, which is optimal for calibrating the longest baselines.
Five targets (3C\,196, 3C\,225, 3C\,273, 3C\,295, and 3C\,298) were observed with the ILT between 12 February and 25 April 2020 for project LC13\_017.
The five targets were observed with the  LBA and used all 13 international stations available at the time.
The \texttt{LBA\_OUTER} configuration was employed, which selects the outer $48$ of $96$ antennas in each Dutch station. Each target was observed for 8~hr, except for 3C\,273, which was observed for only 4 hr due to its low declination.
Data were recorded in the range 10--80~MHz. The data at frequencies below 25~MHz and above 79~MHz were discarded due to the lower bandpass response of the stations. For each source, half the bandwidth in alternating sub-bands was used to simultaneously observe 3C\,196 as a calibrator source. 
The calibrator source was not used throughout the data reduction and was therefore discarded.

3C\,380 was observed with the ILT in project LC15\_024 on 17 and 18 December 2020, in two separate 4~hr observing blocks.
For this observation we used only 12 out of the  13 available international stations as the station in Nan\cb{c}ay (FR606LBA) was not functioning properly. 
We used the \texttt{LBA\_OUTER} dipole selection, but with a different frequency setup: all sub-bands between 22 and 69~MHz were recorded.

Before calibrating the data, the signals from the bright off-axis sources Cassiopeia~A and Cygnus~A were subtracted by demixing the data \citep{2007ITSP...55.4497V,gasperin_ateam}.
In addition, RFI was flagged with \texttt{AOflagger}  \citep{2010ascl.soft10017O,2012A&A...539A..95O}
and the data were compressed with the Dysco software  \citep[][]{2016A&A...595A..99O}.

Calibration was performed in three steps.
The first calibration step consisted of extracting the instrumental effects of the LOFAR stations using data from 3C\,196, our calibrator. We used a high-resolution model of 3C\,196 (Offringa, priv. comm.) and the flux density scale of \cite{2012MNRAS.423L..30S}. The bandpass responses, and instrumental delays between the XX and YY correlations were extracted using \texttt{prefactor} \footnote{\url{https://github.com/lofar-astron/prefactor}} \citep{2019A&A...622A...5D,2016ApJS..223....2V,2016MNRAS.460.2385W}, \texttt{DPPP} \footnote{\url{https://github.com/lofar-astron/DP3}} \citep{2018ascl.soft04003V}, and \texttt{LoSoTo} \footnote{\url{https://github.com/revoltek/losoto}} \citep{2019A&A...622A...5D}.
The corrections that \texttt{prefactor} determines are the bandpass response, and the instrumental delay between the XX and YY correlations \citep[for more details on the polarization alignment calibration, see][]{2019A&A...622A...5D}.
The derived corrections were applied to all six observed sources, including the calibrator.

In the next step we transformed the data from a linear basis to a circular basis for each source.
The reason for this is that perturbations caused by Faraday rotation are represented by only on-diagonal phase terms in a circular basis, which in turn means that we only need to calibrate on-diagonal terms \citep{smirnov_2011_ii}.

Next, we established gain corrections for the core stations by calibrating the Dutch baselines against a simple model from \texttt{prefactor}, using \texttt{DPPP}.
If a \texttt{prefactor} model was not available, we used a model extracted from \citep{2017A&A...601A..35P} (for 3C\,273) or a self-made model (for 3C\,225 and 3C\,298).
We corrected the phases of the core stations with these phase solutions and subsequently combined the data from the core stations into one superstation, reducing the radius of the field of view to $\sim 9$ arcminutes at 50~MHz and increasing the signal-to-noise ratio  of the baselines connecting to the core.

Finally, we self-calibrated the data in order to improve the quality of the image.
In each self-calibration iteration, we calibrated the data against a model, and imaged the data to solve for both clock drift and ionospheric effects simultaneously, similarly to the approach used in the LOFAR LBA Sky Survey \citep[LoLSS,][]{lolss}.
The data was divided into time and frequency bins.
The widths of these bins were determined such that at the lowest frequencies, where the dispersive effects of the ionosphere are the strongest, the differences in phase solution do not exceed 1 radian between two adjacent bins.
Calibration was done using \texttt{DPPP}, and imaging with \texttt{WSClean} \citep{2014MNRAS.444..606O,2017MNRAS.471..301O}.
This produces images with a beam size between 0.42 arcsecond and 1.1 arcsecond, depending on the declination of the target.
After the initial imaging the resulting model was used as starting model for the next calibration cycle.
During the first three cycles of self-calibration, we restricted the calibration to phase-only solutions, which prevents \texttt{DPPP} from diverging.
Subsequently, we allowed the complex gains (amplitude and phases) to be calibrated after applying the phase-only calibration.
After three cycles of amplitude and phase calibration, the calibration results were stabilized and subsequent calibration cycles did not show an appreciable improvement in the final image.

For all imaging, we used Briggs weighting with a robustness parameter of $-1.5$ and a $uv$-cut of 500 $\lambda$ (corresponding to 2.5 kilometers at 60\,MHz).
Cleaning was restricted to regions that were at least 5$\sigma$ above the noise (automasking).
Manual masks did not significantly improve the quality of the final image, so we relied on applying automasking.
After imaging, the cleaned model was clipped to remove negative emission (during amplitude+phase calibration).

\begin{table*}
\centering
\caption{Fluxes measured with LOFAR LBA of each source. These fluxes correspond to the filled diamonds in \autoref{fig:comp_cats}.}
\begin{tabular}{lr}
    \multicolumn{2}{c}{\textbf{\large 3C196}} \\
    \textbf{$\nu$ (MHz)} & \textbf{$S$ (Jy)} \\ \hline
    27 - 38 & 187 $\pm$ 48\\
    39 - 49 & 166 $\pm$ 44\\
    49 - 59 & 152 $\pm$ 41\\
    59 - 69 & 141 $\pm$ 39\\
    69 - 79 & 130 $\pm$ 36
\end{tabular}
\hspace{1cm}
\begin{tabular}{lr}
    \multicolumn{2}{c}{\textbf{\large 3C225}} \\
    \textbf{$\nu$ (MHz)} & \textbf{$S$ (Jy)} \\ \hline
    29 - 39 & 47.4 $\pm$ 21\\
    39 - 49 & 45.1 $\pm$ 21\\
    49 - 58 & 41.4 $\pm$ 20\\
    58 - 68 & 47.8 $\pm$ 21\\
    68 - 78 & 38.4 $\pm$ 19
\end{tabular}
\hspace{1cm}
\begin{tabular}{lr}
    \multicolumn{2}{c}{\textbf{\large 3C273}} \\
    \textbf{$\nu$ (MHz)} & \textbf{$S$ (Jy)} \\ \hline
    29 - 39 & 327 $\pm$ 89\\
    39 - 49 & 216 $\pm$ 67\\
    49 - 58 & 186 $\pm$ 61\\
    58 - 68 & 175 $\pm$ 59\\
    68 - 78 & 180 $\pm$ 60
\end{tabular}\\
\vspace{1cm} \hspace{1cm}
\begin{tabular}{lr}
    \multicolumn{2}{c}{\textbf{\large 3C295}} \\
    \textbf{$\nu$ (MHz)} & \textbf{$S$ (Jy)} \\ \hline
    25 - 35 & 93.2 $\pm$ 24\\
    35 - 46 & 115 $\pm$ 28\\
    46 - 56 & 125 $\pm$ 30\\
    56 - 67 & 132 $\pm$ 32\\
    67 - 78 & 120 $\pm$ 29
\end{tabular}
\hspace{1cm}
\begin{tabular}{lr}
    \multicolumn{2}{c}{\textbf{\large 3C298}} \\
    \textbf{$\nu$ (MHz)} & \textbf{$S$ (Jy)} \\ \hline
    29 - 39 & 75 $\pm$ 21\\
    39 - 49 & 80.1 $\pm$ 22\\
    49 - 58 & 100 $\pm$ 26\\
    58 - 68 & 111 $\pm$ 28\\
    68 - 78 & 103 $\pm$ 26
\end{tabular}
\hspace{1cm}
\begin{tabular}{lr}
    \multicolumn{2}{c}{\textbf{\large 3C380}} \\
    \textbf{$\nu$ (MHz)} & \textbf{$S$ (Jy)} \\ \hline
    24 - 35 & 222 $\pm$ 64\\
    37 - 45 & 183 $\pm$ 56\\
    46 - 54 & 169 $\pm$ 54\\
    54 - 62 & 161 $\pm$ 52\\
    62 - 71 & 143 $\pm$ 48
\end{tabular}
\hspace{1cm}
\label{tab:fluxes_lofar}
\end{table*}

\begin{figure}
    \hspace{-0.75cm}
    \includegraphics[width=1.1\linewidth]{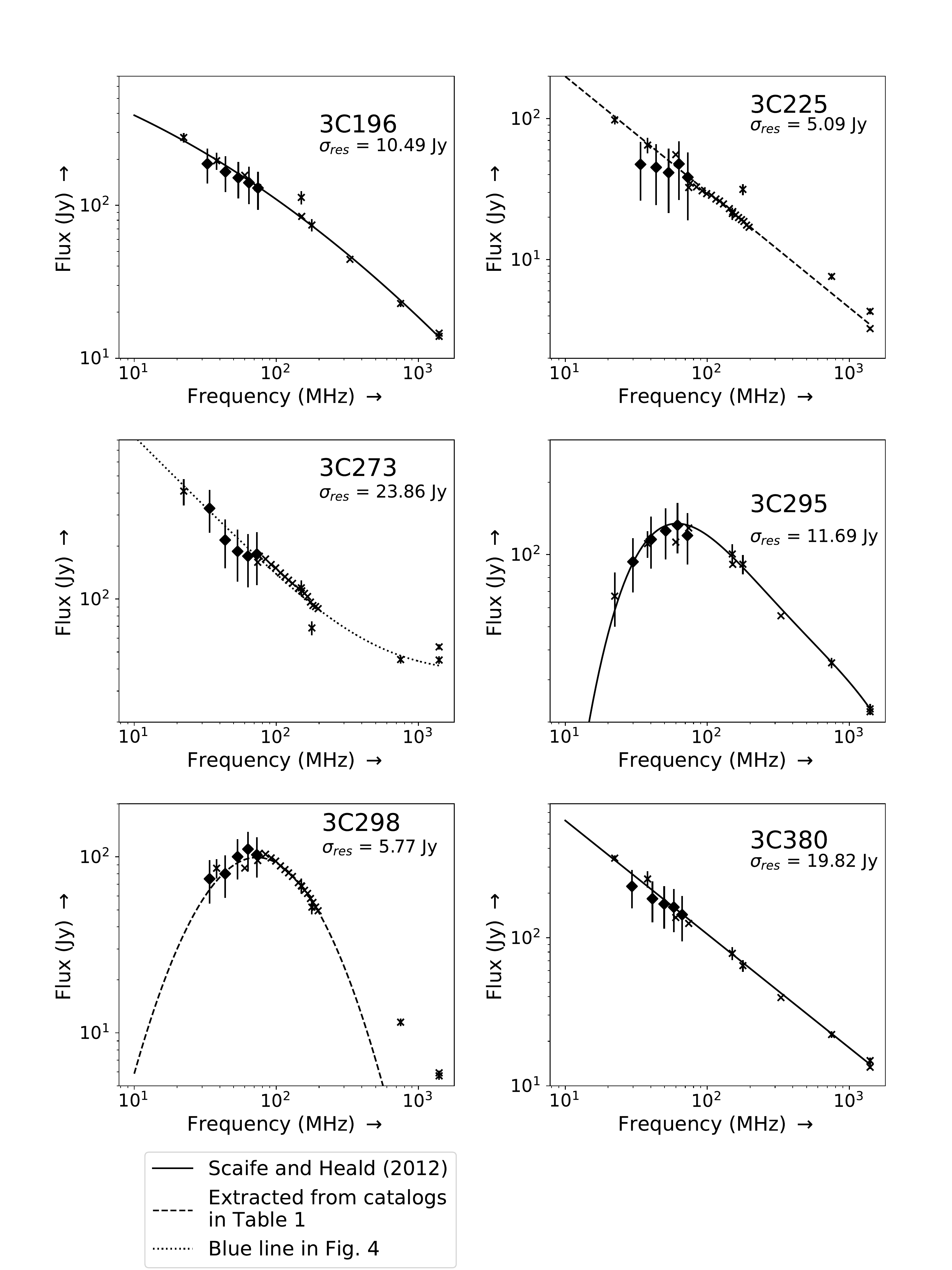}
    \caption{Low-frequency spectra for our targets. The low-frequency flux densities were measured using the Dutch LOFAR array (see also \autoref{tab:fluxes_lofar}). The filled diamonds represent LOFAR flux densities, and the crosses represent data extracted from the catalogs listed in \autoref{tab:catref}. For 3C\,196, 3C\,295, and 3C\,380 the black solid line shows the flux scale model of \cite{2012MNRAS.423L..30S}. For 3C\,273 the fitting result of the blue line (total flux) of \autoref{fig:ap_perleylofar} is shown, with the functional form of \autoref{eq: 3c273_model}. The differences between the flux densities and the fit (residuals) are approximated by a Gaussian distribution, and the standard deviations of these residuals are given for each target in the figure. The error bars on the LOFAR data consist of 20\% of the flux density added to the standard deviation of the residuals. The uncertainties of the other data points were extracted from the literature.}
    \label{fig:comp_cats}
\end{figure}

We checked the flux density scale by making images solely with the Dutch LOFAR baselines for which the sources are unresolved. This also avoids potential issues with phase decorrelation from imperfect phase calibration on the international baselines.
Phase decorrelation will lead to loss of flux density, especially on long baselines at low frequencies, due to the very rapid ionospheric phase variations as a function of frequency and time.
For this reason we also do not give any absolute flux measurements below 30 MHz.
The measured flux densities are presented in \autoref{tab:fluxes_lofar} and \autoref{fig:comp_cats}, along with flux densities from other catalogs mentioned in \autoref{tab:catref}.
For 3C\,196, 3C\,295, and 3C\,380 we show the flux density models from \cite{2012MNRAS.423L..30S} as solid lines. Our measured flux densities are in good agreement with \citeauthor{2012MNRAS.423L..30S} for these three sources, demonstrating that during our self-calibration the flux density scale is well preserved.
For the other sources, we  plot the models fitted to the flux densities extracted from the catalogs listed in \autoref{tab:catref}.
For 3C\,225 we fit a power-law model, and for 3C\,298 a second-order polynomial in log-log space. The fit for 3C\,273 is described in Sect.~\ref{sec:3C273}. Comparing the LOFAR flux densities to those in  the catalogs we again find consistent results. Based on these comparisons, and the more detailed analysis carried out on LBA data by \cite{lolss}, we adopt a conservative absolute flux density scale uncertainty of 20\%.

\begin{table}[ht]
    \centering
    \caption{Catalogs used in \autoref{fig:comp_cats} to retrieve the flux densities of the sources. The flux densities extracted from these catalogs were corrected to be on the same flux scale as those in \cite{2012MNRAS.423L..30S}.}
    \begin{tabular}{l r r}
    Name & \begin{tabular}{r}Frequencies\\(MHz)\end{tabular} & Reference  \\ \hline
    3CR      & 10--1400& \cite{1980MNRAS.190..903L} \\
             & 38      & \cite{1969ApJ...157....1K} \\
    DRAO     & 22      & \cite{1986AAS...65..485R}  \\
    VLSSR    & 74      & \cite{2014MNRAS.440..327L} \\
    GLEAM    & 76--227 & \cite{2017MNRAS.464.1146H} \\
    AARTFAAC & 60      & \cite{aartfaac}            \\
    GMRT     & 150     & \cite{tgssadr} \\
    7C       & 151     & \cite{2007MNRAS.382.1639H} \\
    WENSS    & 330     & \cite{2000yCat.8062....0D} \\
    NVSS     & 1400    & \cite{nvss}                \\
    \end{tabular}
    \label{tab:catref}
\end{table}

\begin{figure*}[]
    \centering
    \includegraphics[width=0.45\linewidth]{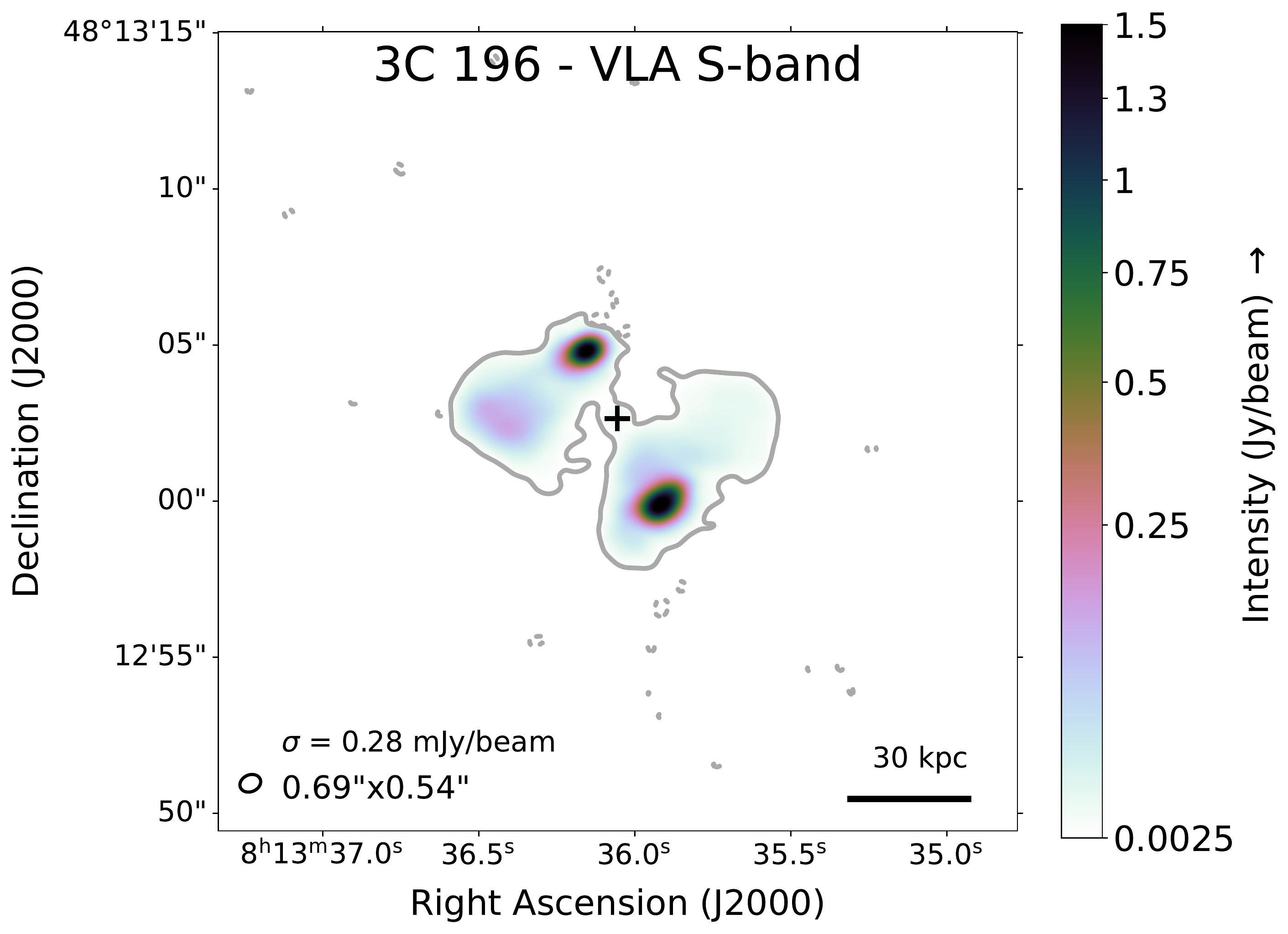}
    \includegraphics[width=0.45\linewidth]{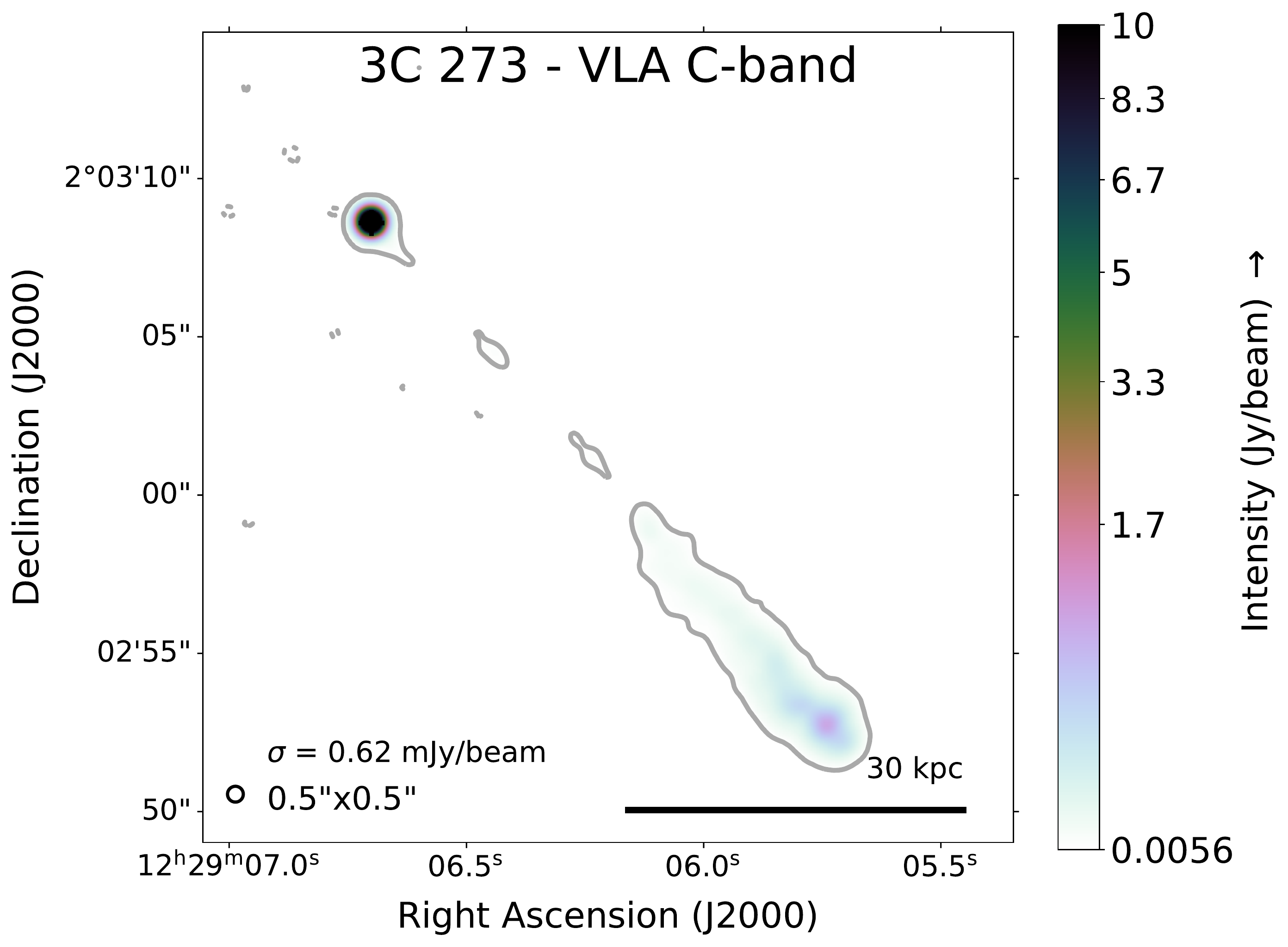}
    \caption{VLA images of 3C\,196 and 3C\,273. Left:  2.5~GHz VLA S-band image of 3C\,196  obtained from project \texttt{TCAL0001}. Right:  4.9~GHz image of 3C\,273, reproduced from \citep{2017A&A...601A..35P}. The cross indicates the optical counterpart, obtained from Gaia EDR3 \citep{gaiaedr3}. The brightness scale starts at the 9$\sigma$ noise level ($\sigma_{\rm{rms}}$), which is also indicated by the single solid contour. }
    \label{fig:VLA_3c196}
\end{figure*}

Our initial self-calibration models were simplistic and poorly represented the actual structure. In addition, positions are not preserved under self-calibration. For these reasons the astrometric accuracy of our images could be compromised. We verified the astrometric accuracy by overlaying the optical host positions and by comparing them to higher frequency maps from the VLA. For three sources, 3C\,273, 3C\,225, and 3C\,298, we found that a position shift of a few arcseconds was required to properly align the LBA maps.

\subsection{VLA data of 3C\,196 and 3C\,273}
For 3C\,196, we used additional VLA S-band data (frequency range 1.93--3.14~GHz), recorded on 12 December 2016 with project code \texttt{TCAL0001}, to generate a spectral index map.
The data were reduced using CASA.
The data were Hanning smoothed, before the shadowed antennas were flagged, gain elevation curves applied, and antenna positions corrected.
The RFI  was excised using \texttt{AOflagger} \citep{2010ascl.soft10017O,2012A&A...539A..95O}.
Using the bandpass calibrator 3C\,286, the initial complex gain solutions and the delays were computed, and the bandpass calibration solutions were determined.
The flux scale used was obtained from \cite{perleybutler2017}.
This flux scale is accurate from 50~MHz to 50~GHz, and is different from the \citet{2012MNRAS.423L..30S} flux scale used in the LOFAR data calibration.
At a frequency of 73.8~MHz the Perley-Butler flux density of 3C\,196 is 98.1\% of the Scaife-Heald flux density scale, and at 50~MHz the Perley-Butler flux density is 103\% of the Scaife-Heald flux density scale.
This offset is lower than the total flux density error of the LOFAR images.
This translates to an error of 0.84\% on the integrated spectral index of 3C\,196.
In addition, this error only impacts the spectral index of the integrated spectra of the sources, and not the relative spectral indices in resolved maps.

After applying the previous calibration solutions, we re-computed the complex gain solutions using the phase calibrator J0808+4950, and we applied the final complex gain solution.
The result is imaged with \texttt{tclean}.
We applied four cycles of additional self-calibration to refine the calibration solutions, and the final result is shown in \autoref{fig:VLA_3c196}.

For 3C\,273, we used the images provided in \cite{2017A&A...601A..35P}.
A 3.9~GHz C-band image is reproduced in \autoref{fig:VLA_3c196}.

\section{Results}
\label{sec:results}

\begin{figure*}[!ht]
    \centering
  \resizebox{\textwidth}{!}{
    \includegraphics[width=0.3\textwidth,clip]{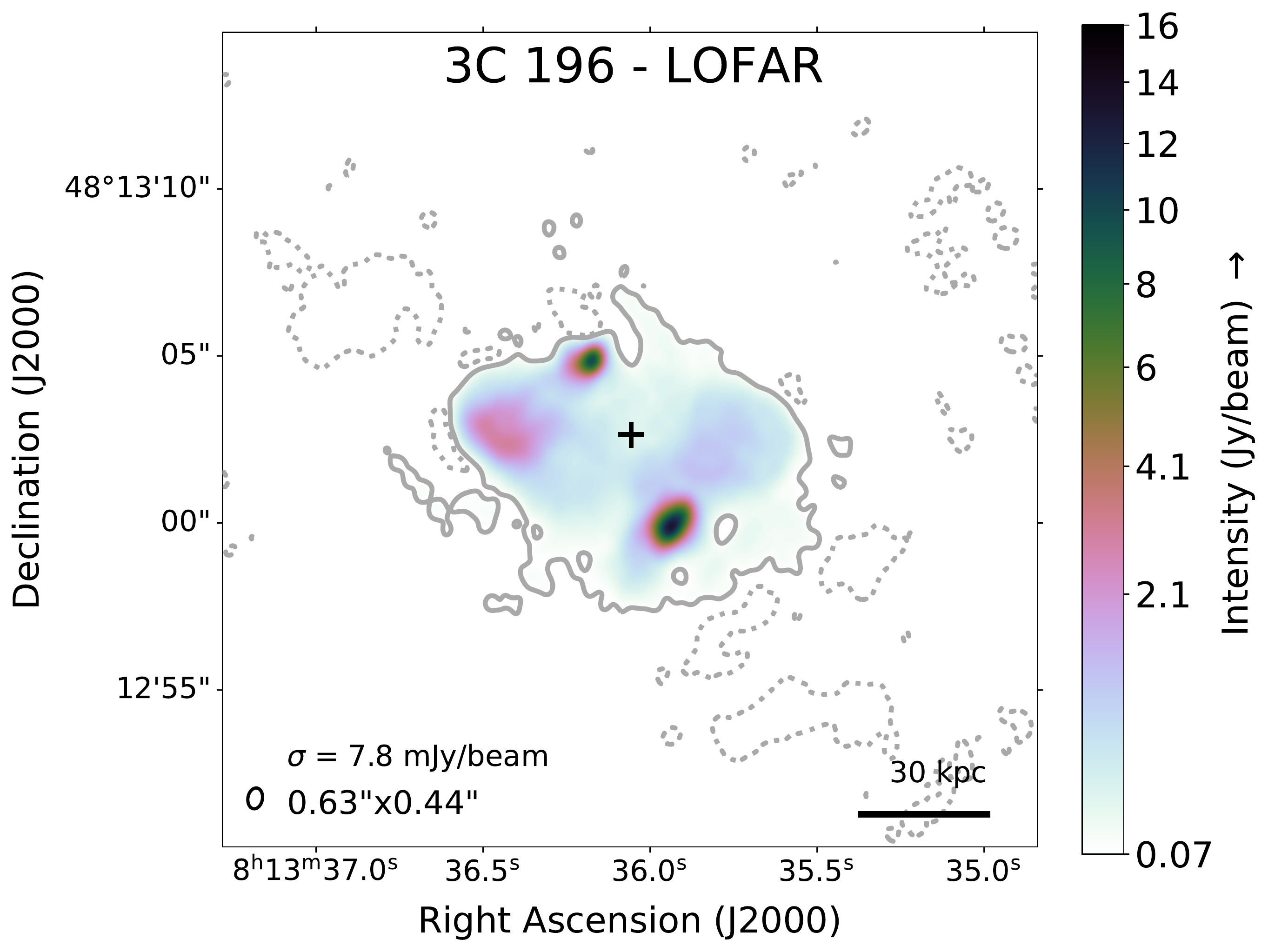}
    \includegraphics[width=0.3\textwidth,clip]{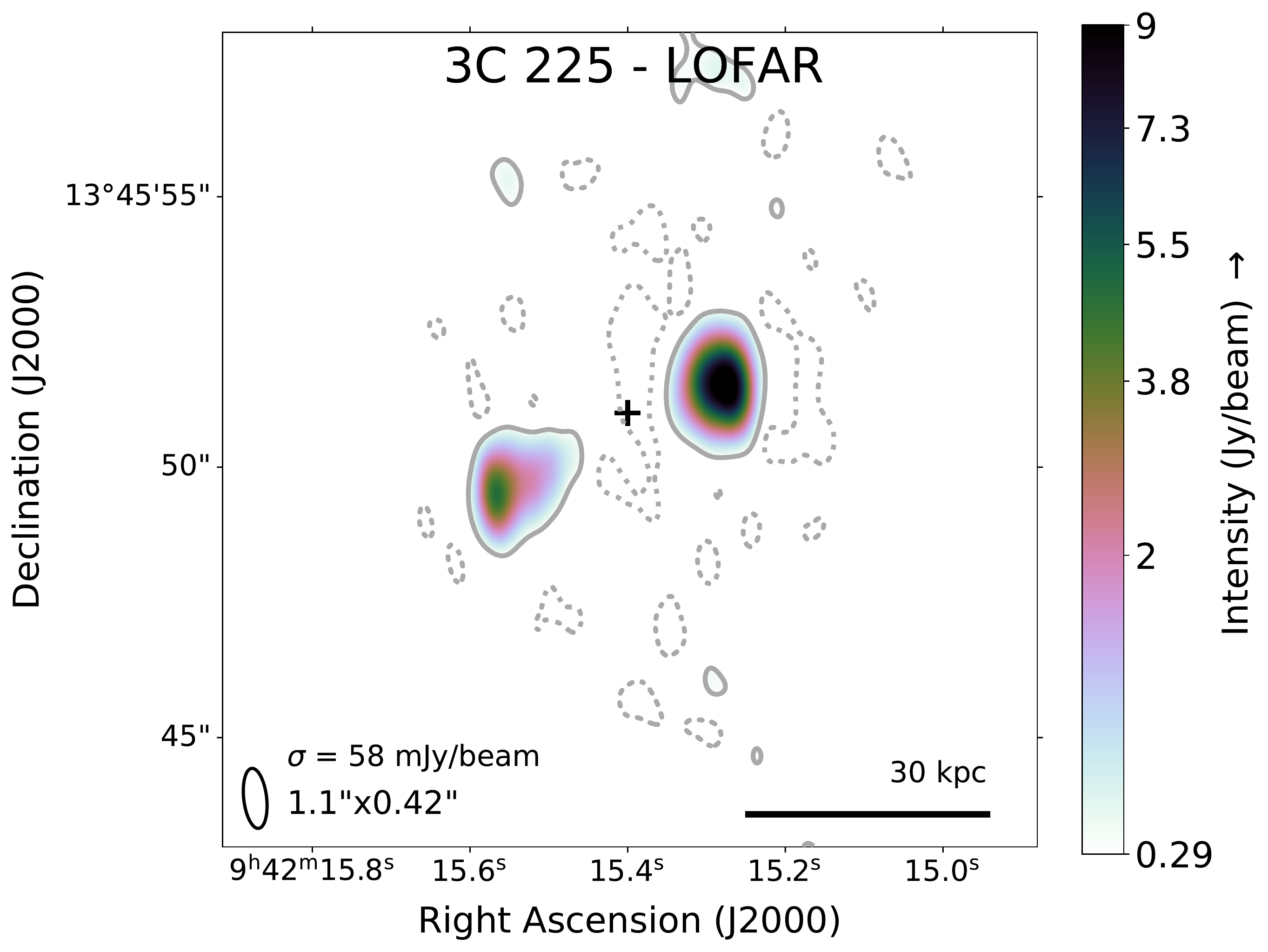}
    \includegraphics[width=0.3\textwidth,clip]{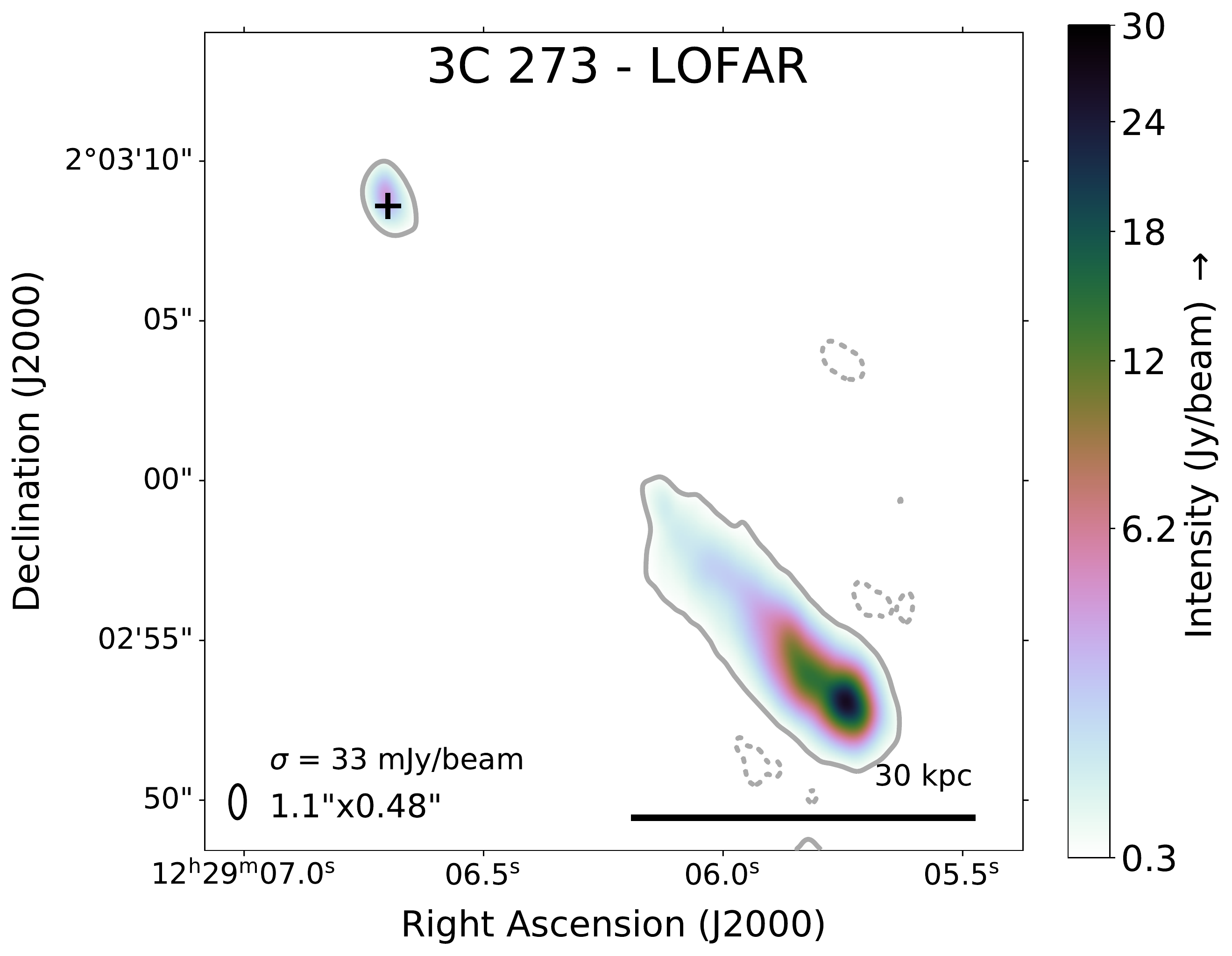}
  }\vspace{12pt}
  \resizebox{\textwidth}{!}{
    \includegraphics[width=0.3\textwidth,clip]{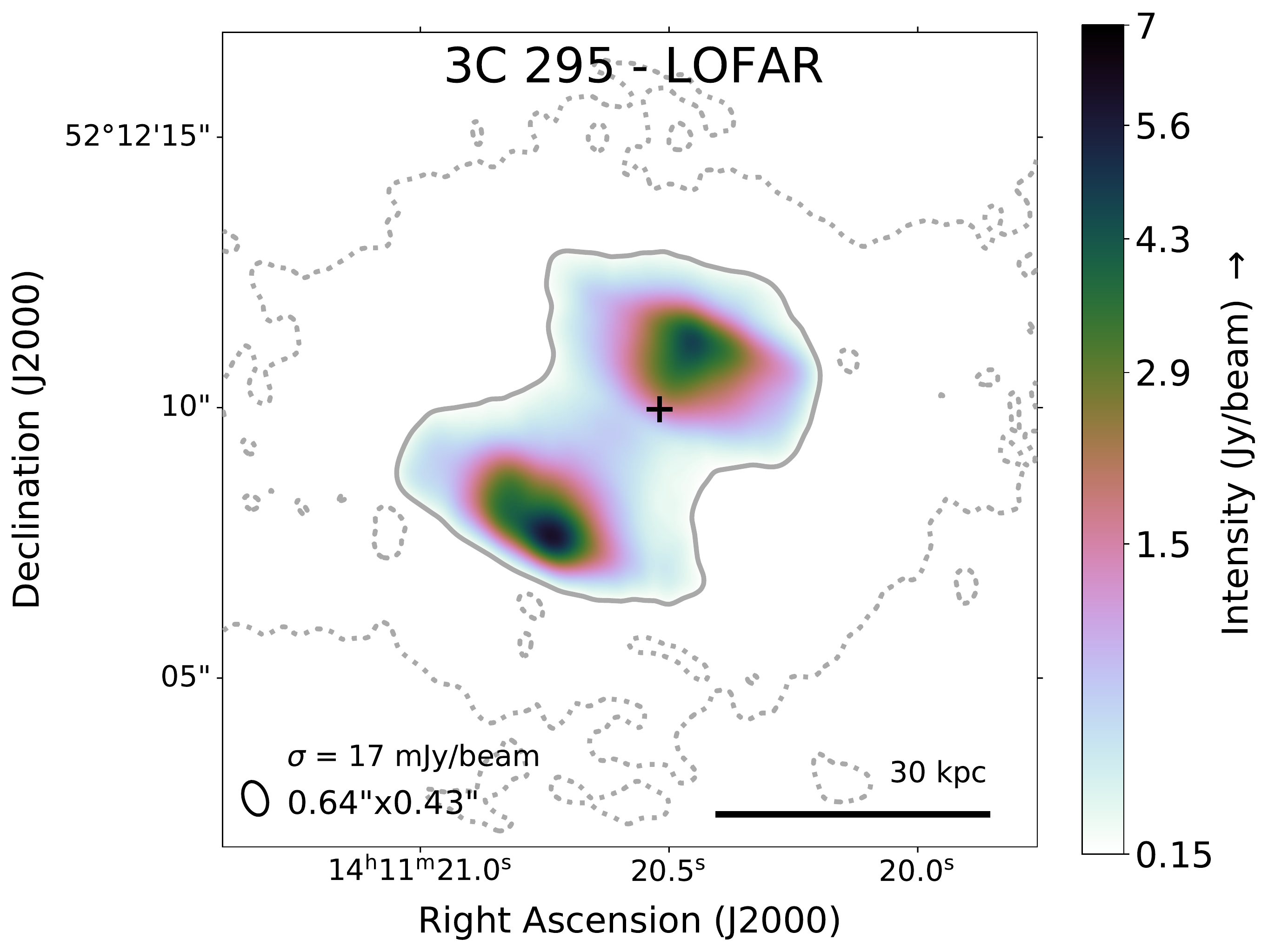}
    \includegraphics[width=0.3\textwidth,clip]{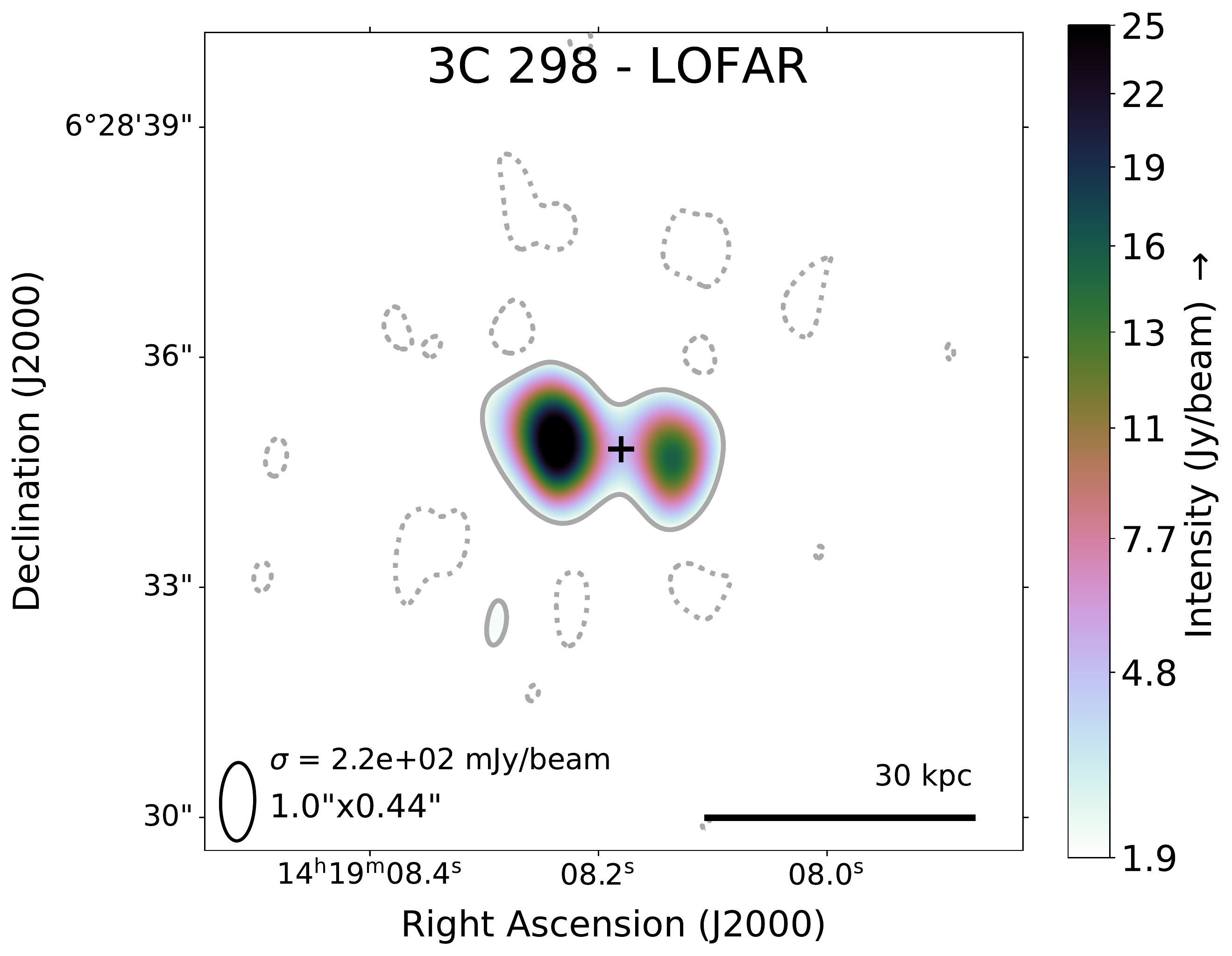}
    \includegraphics[width=0.3\textwidth,clip]{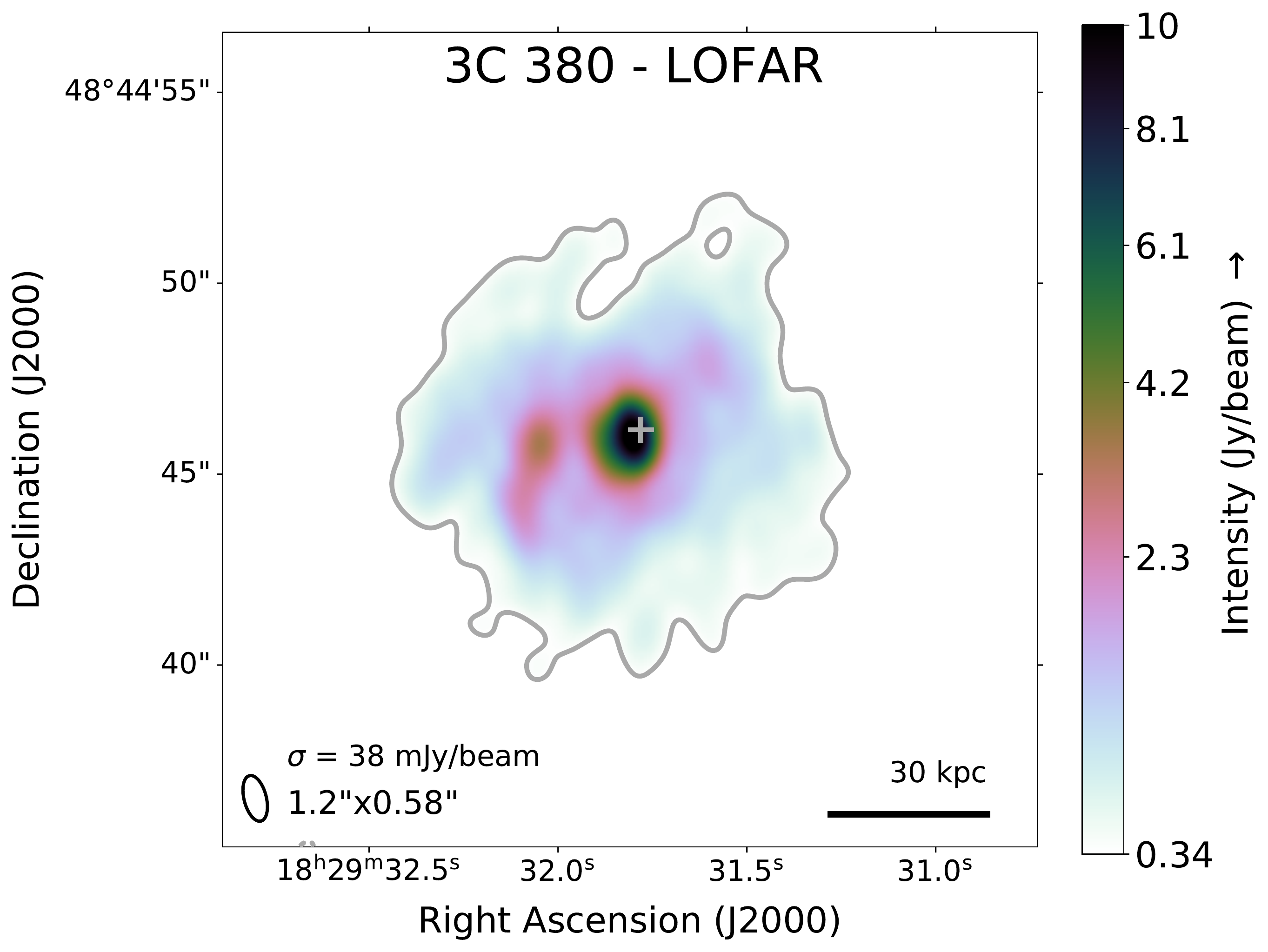}
  }
  \caption{Images of the six  3C sources, observed in the range 30--80 MHz. The beam sizes are indicated in the bottom left corner. The optical counterparts of the sources are indicated with crosses. The positions of the optical counterparts were taken from  Gaia EDR3 \citep{gaiaedr3} (3C\,196, 3C\,298, and 3C\,273), \cite{ReyesSDSS} (3C\,295), and \cite{FantiCSSVLA} (3C\,273).
  The brightness scale starts at the 9$\sigma$ noise level ($\sigma_{\rm{rms}}$), which is also indicated by the single solid contour. The dashed contours reveal spurious negative emission due to calibration artifacts at the $-3 \sigma_{\rm{rms}}$ level.}
  \label{fig:full_plots}
  \centering
  \resizebox{\textwidth}{!}{
    \includegraphics[width=0.3\textwidth,clip]{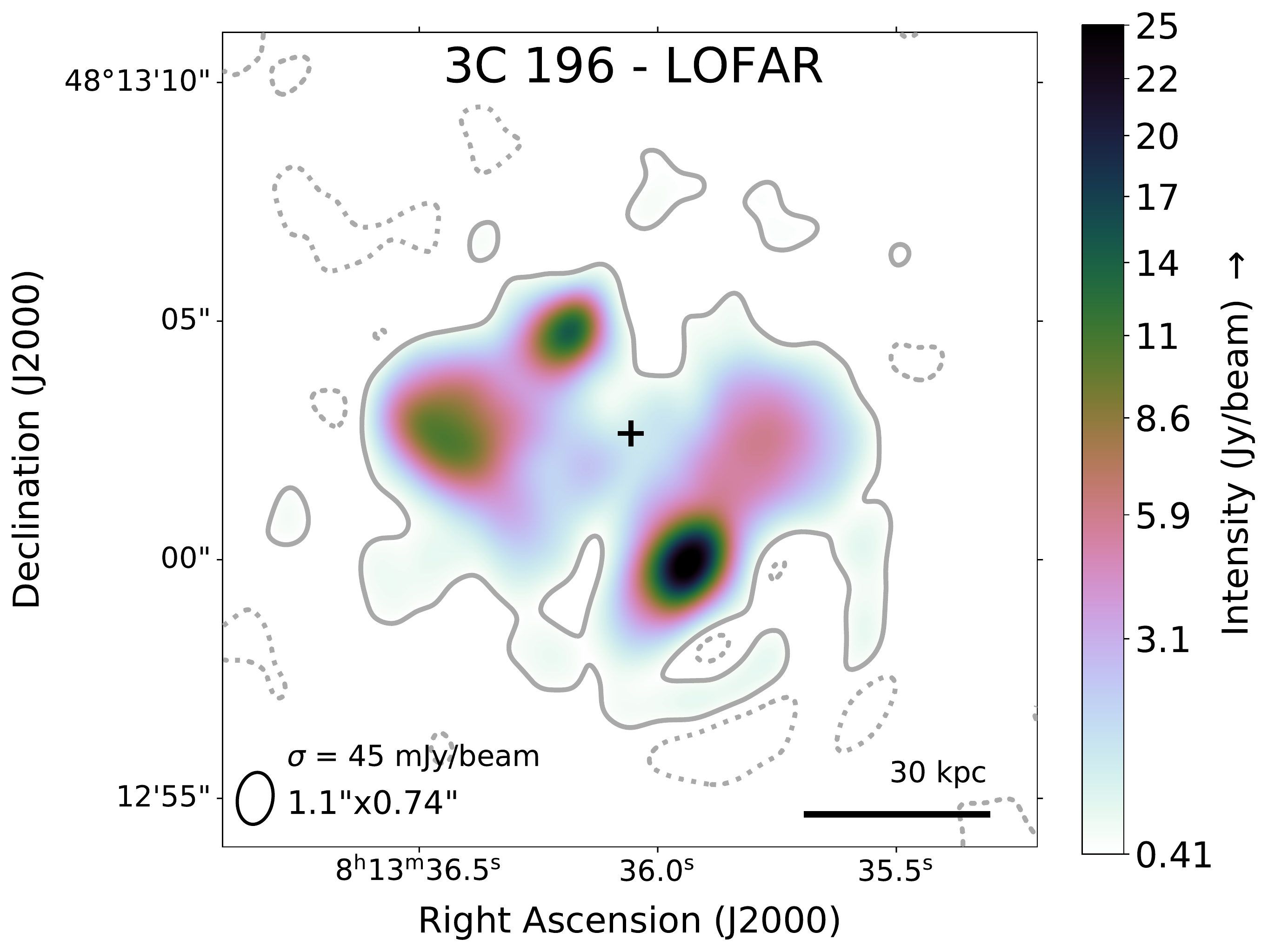}
    \includegraphics[width=0.3\textwidth,clip]{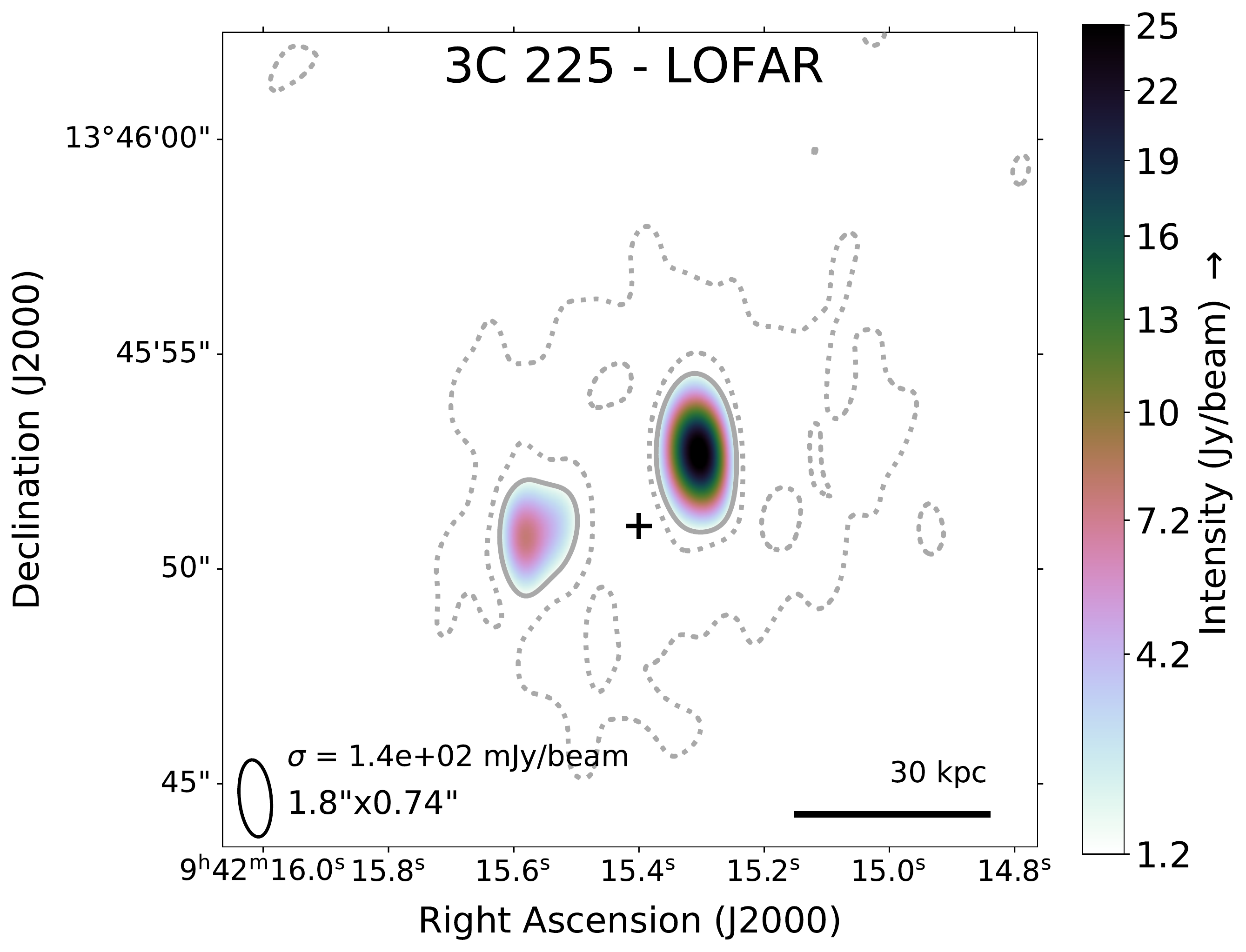}
    \includegraphics[width=0.3\textwidth,clip]{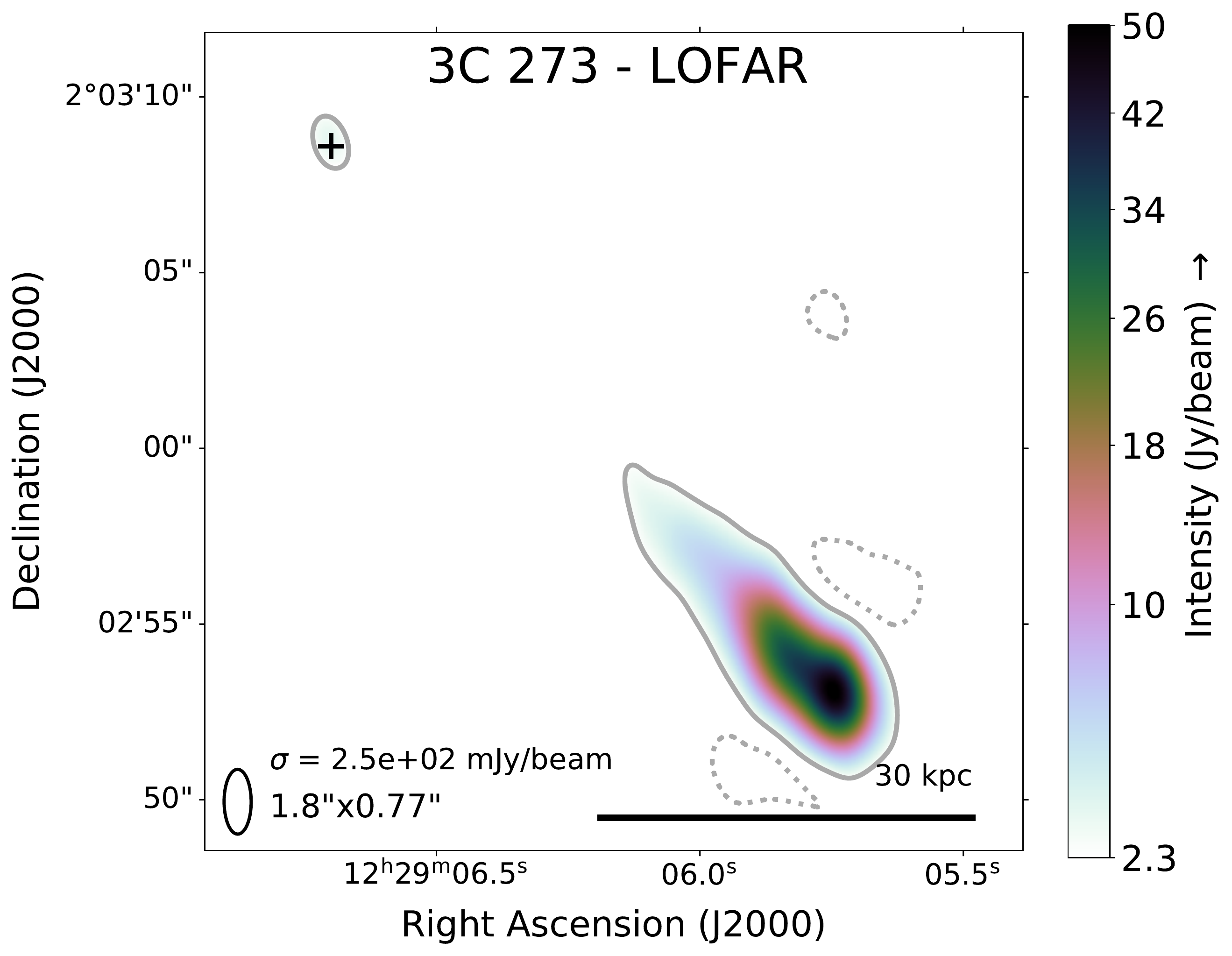}
  }\vspace{12pt}
  \resizebox{\textwidth}{!}{
    \includegraphics[width=0.3\textwidth,clip]{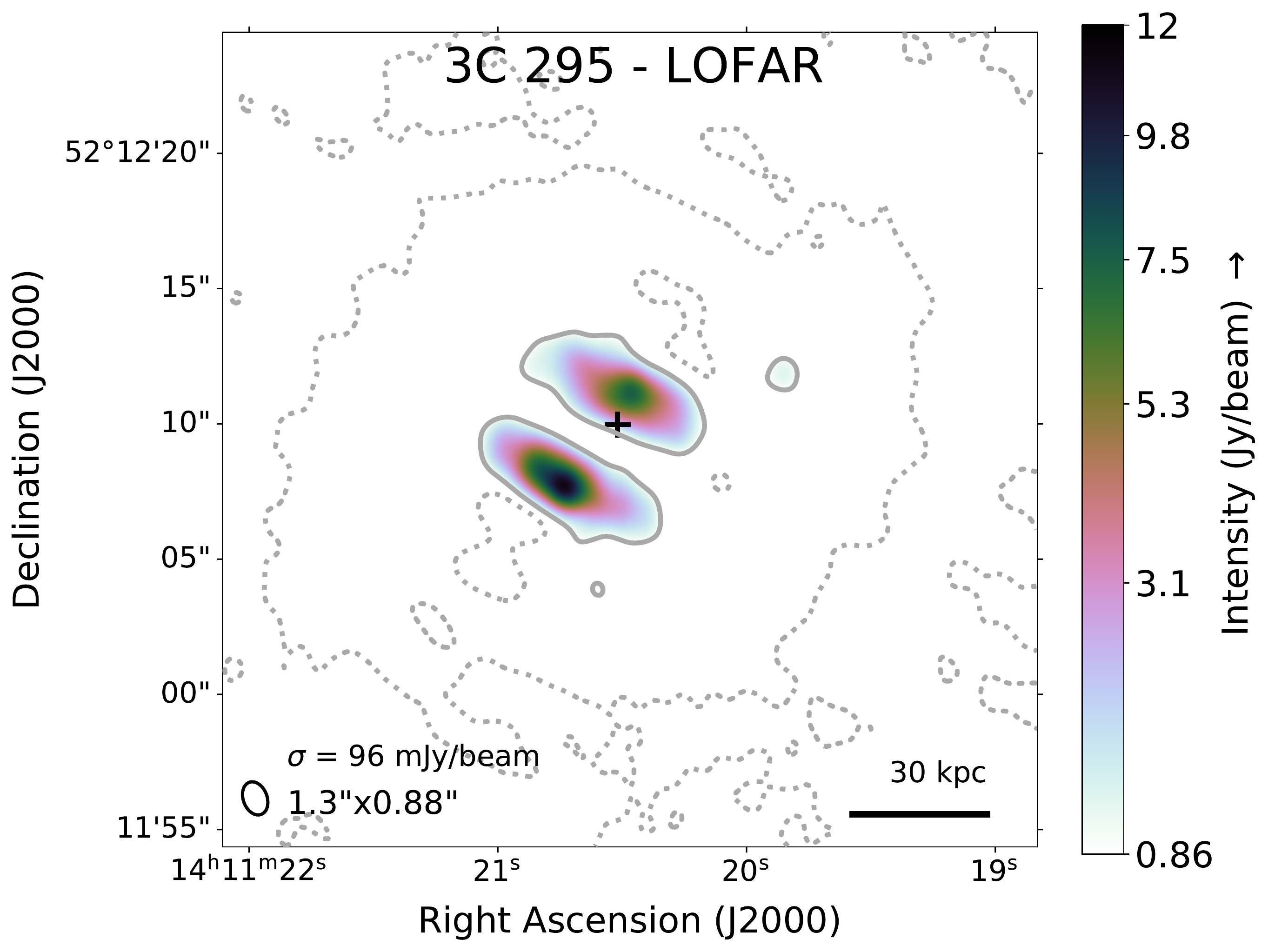}
    \includegraphics[width=0.3\textwidth,clip]{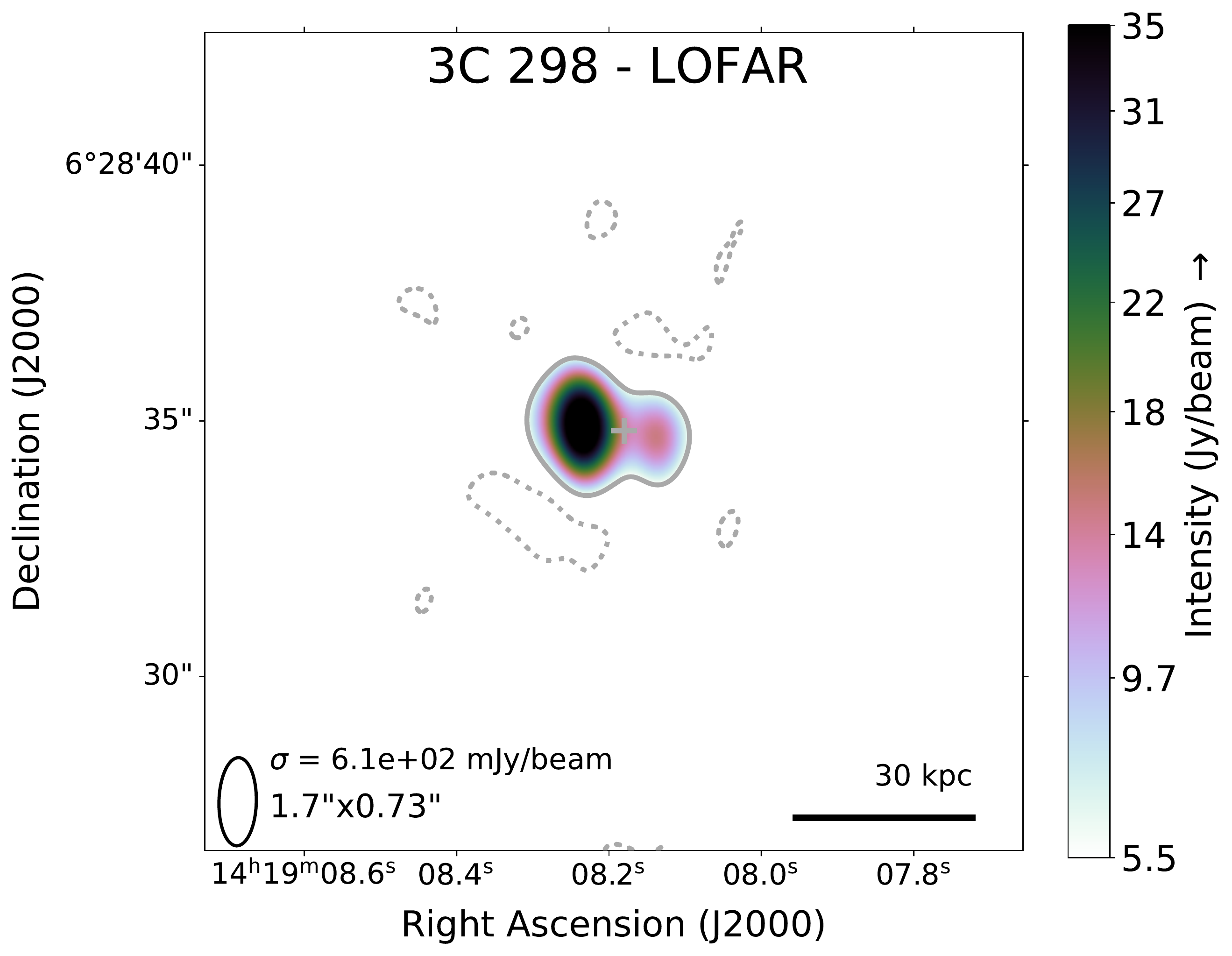}
    \includegraphics[width=0.3\textwidth,clip]{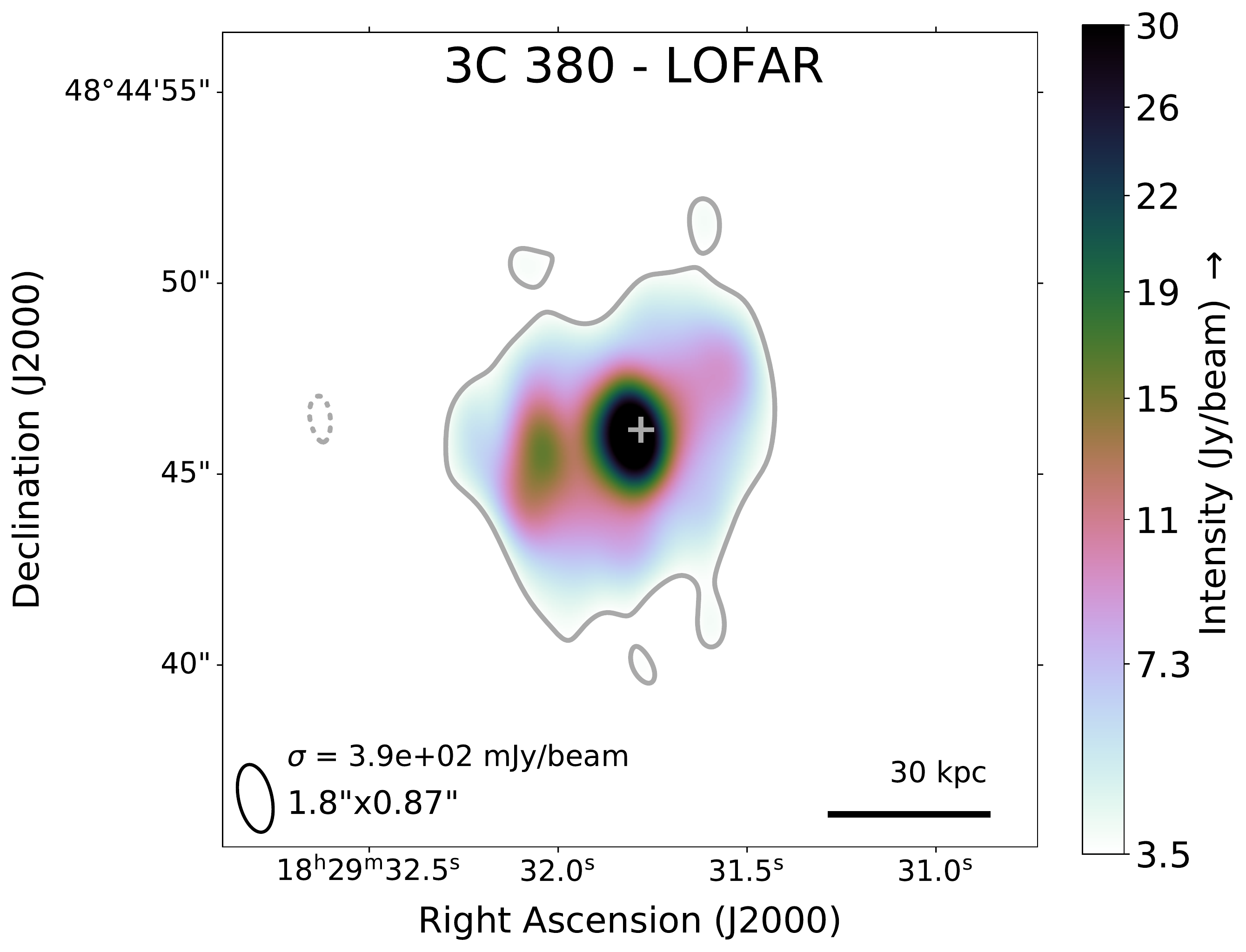}
  }
  \caption{Images of the six targets, observed in the  30--40~MHz ULF range. The  flux scales were set with Dutch LOFAR. These images are similar to \autoref{fig:full_plots}, but limited to frequencies below 40 MHz. The quality of the images is lower than the full broadband images, but they reveal the structure at low frequencies. The images have more than an order of magnitude higher resolution than pre-ILT images at ULF.}
  \label{fig:lowfreq}
\end{figure*}

For all six of our sources, we successfully produced a subarcsecond resolution image, achieving a dynamic range of at least 140 for the full bandwidth and a dynamic range of at least 70 between 30 and 40~MHz. The full bandwidth images are shown in \autoref{fig:full_plots}, with images between 30 and 40~MHz in \autoref{fig:lowfreq}. In the following  we describe the results for these sources individually.

\subsection{3C\,196}
3C\,196 is associated with an 18{th} magnitude quasar \citep{2010A&A...518A..10V} at a redshift of 0.871 \citep{1985PASP...97..932S}.
Radio images at 408~MHz   in \cite{1980Natur.288...66L} show a four-component structure, with two bright and two fainter lobes.
The faint east--west oriented lobes have a steep spectrum, suggesting that these lobes are produced by a relatively old population of electrons and that the source jet axis may have rotated during the last ~20 kiloyears \citep{1980Natur.288...66L}.
Additionally, the existence of a gap between the components suggests that during the jet precession, a period of low AGN activity separated two epochs of high activity.
Such a structure is also known as a ``winged'' or X-shaped radio source \citep{xshaped_wings}.
A similar structure can be obtained when assuming that the eastern and western lobes are caused by backflows from the northern and southern lobes, respectively, where the jet material deposited in the hotspots flows back in another direction \citep[for more details, see, e.g.,][]{antonucci2010_backflows}.
\citet{2010iska.meetE..58W} imaged this source between 30 and 80~MHz with long baselines, although the full ILT was not available at that time (only the Dutch and German stations were available), which limited sensitivity and resolution.

\begin{figure*}
    \centering
    \resizebox{\textwidth}{!}{
        \includegraphics[width=0.46\textwidth,clip]{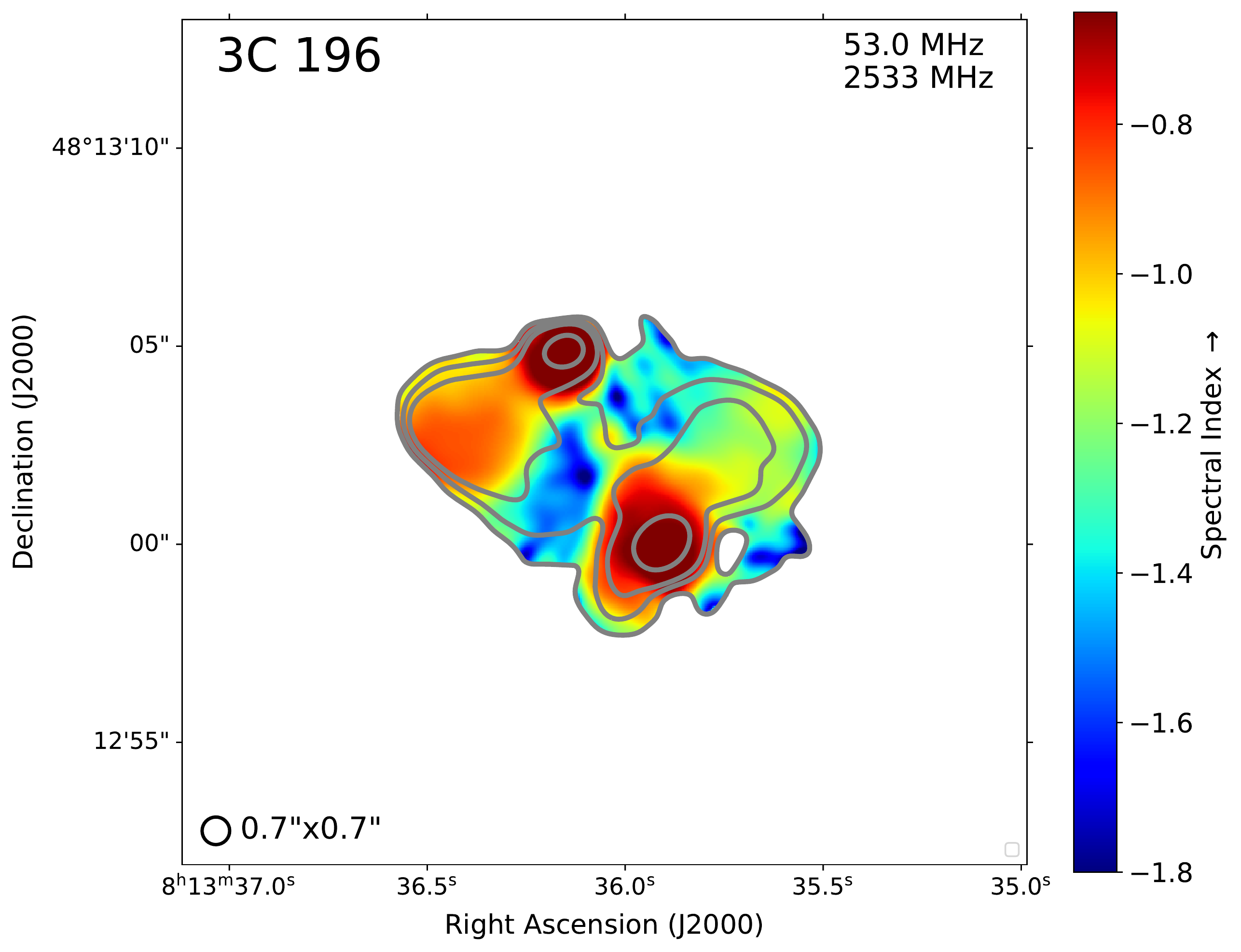}\hspace{1cm}
        \includegraphics[width=0.46\textwidth,clip]{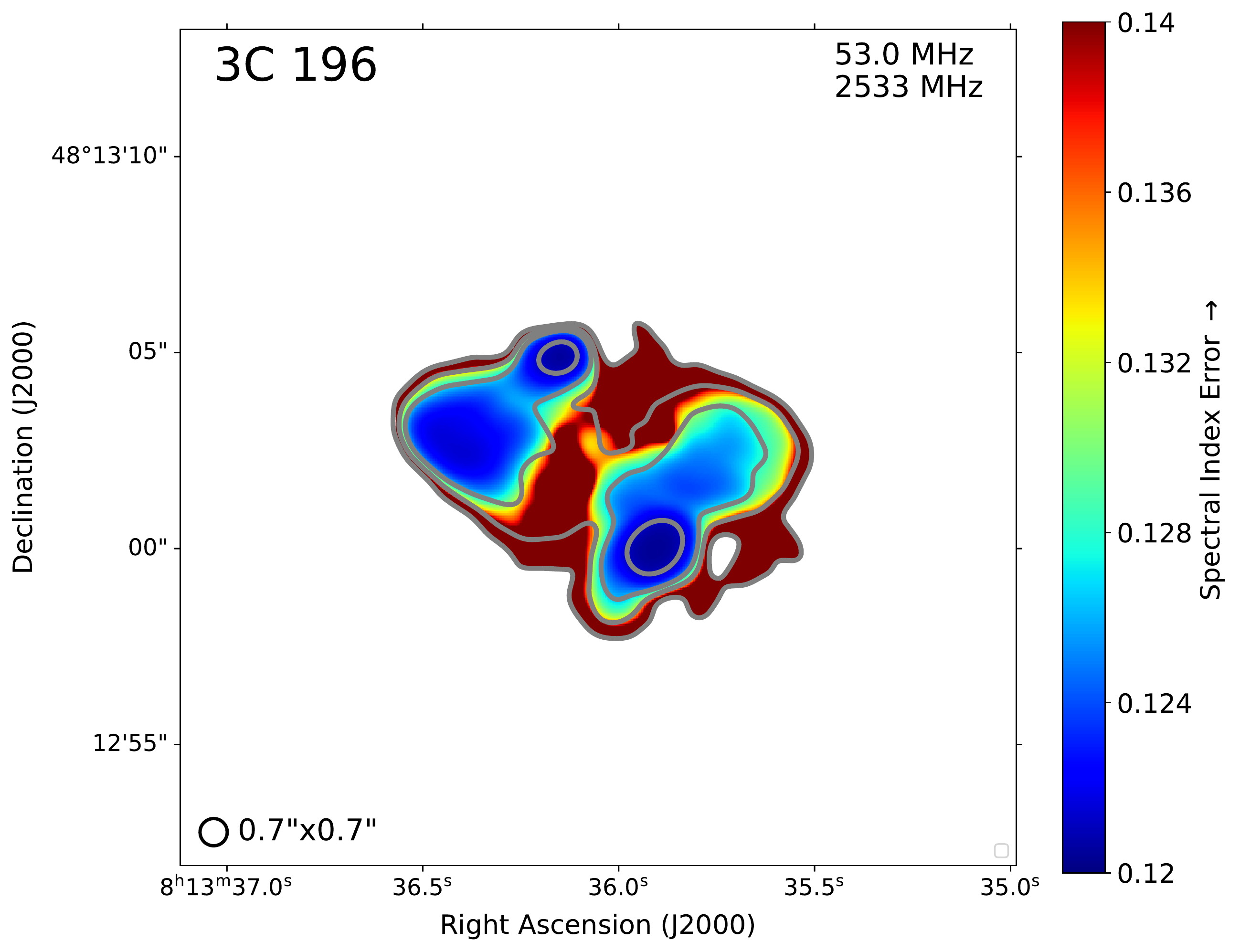}
    }
    \resizebox{\textwidth}{!}{
        \includegraphics[width=0.46\textwidth,clip]{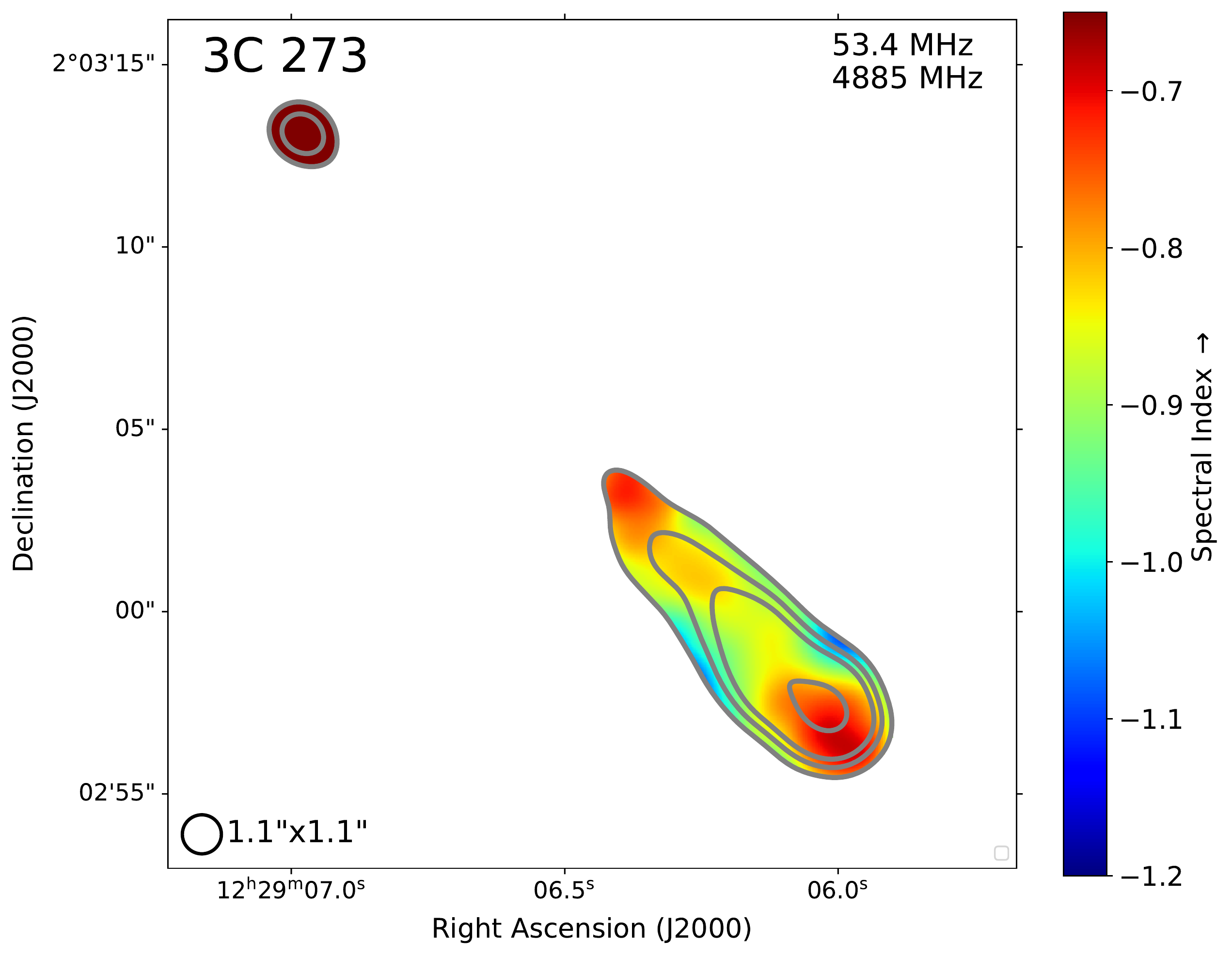}\hspace{1cm}
        \includegraphics[width=0.46\textwidth,clip]{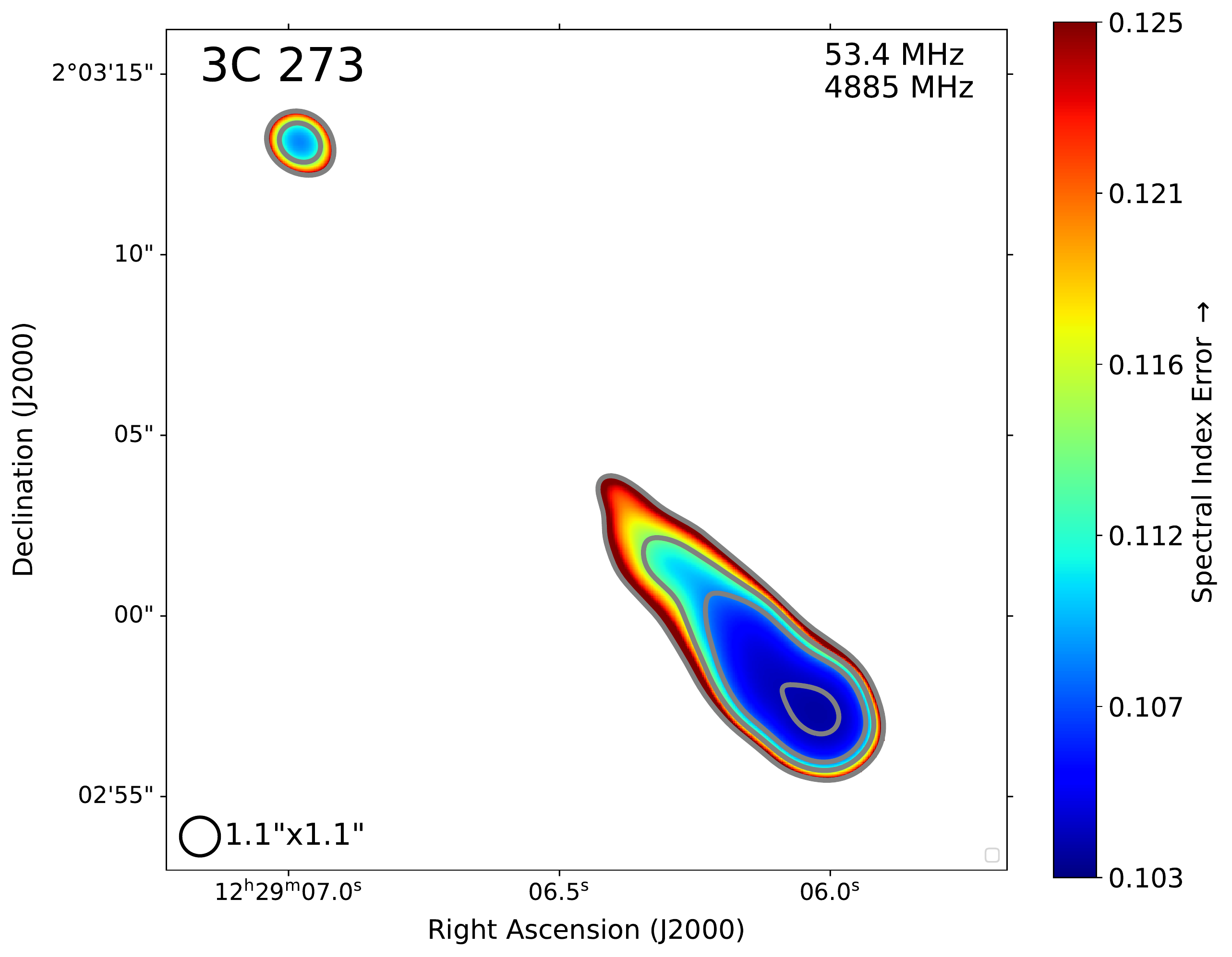}
    }
    \caption{Spectral index maps and spectral index error maps of 3C\,196 and 3C\,273. \textit{Left: }Spectral index maps of 3C\,273 and 3C\,196, between LBA and VLA C-band and VLA S-band, respectively. For each image  the VLA and LBA images are  convolved to a matching circular beam. Contours correspond to LOFAR LBA fluxes, at $20\sigma, 50\sigma, 100\sigma, \text{ and } 500\sigma$, with $\sigma=\SI{0.062}{\jansky}$ for 3C\,273 and $\sigma=\SI{0.011}{\jansky}$ for 3C\,196. These noise levels are different from the noise levels in \autoref{fig:full_plots}, due to the beam being smoothed to a circular beam in order to match the VLA resolution. Data from \citet{2017A&A...601A..35P} were used for 3C\,273, and publicly available VLA data from project \texttt{TCAL0001} for 3C\,196, along with the LBA data presented in this paper. \hspace{\textwidth} \textit{Right: }Spectral index error maps of 3C\,273 and 3C\,196. The errors are made under the assumption that the LOFAR fluxes have a total error consisting of the RMS noise + 20\% flux error. Similarly, this assumes that the VLA fluxes have a total error consisting of the RMS noise + 2.5\% flux error. The error caused by the relative flux error on the VLA and LOFAR fluxes is independent of the position (i.e., irrelevant for comparing two regions of the same source). This error is equal to the minimum value of the color bar.}
    \label{fig:spixes}
\end{figure*}

The eight-hour ILT LBA observation of 3C\,196 results in an image with 0.65~arcsecond resolution and a RMS noise of 8~\mjybeam.
For this source we also show a spectral index map in \autoref{fig:spixes}, combining the LOFAR LBA image with the VLA S-band at 2500~MHz.
The result shows that the two brighter lobes have a synchrotron spectrum with a spectral index of $\sim -0.7$, while the two fainter  lobes have a steeper spectral index of $-0.85$ to $-1.15$.
The core is not visible at LBA frequencies, due to its relatively flat spectrum.
The ULF observations also reveal that the source is embedded in a region of diffuse steep-spectrum emission.
This may be due to aged relativistic plasma produced in previous episodes of AGN activity.

\subsection{3C\,225}
3C\,225B\footnote{3C\,225 has an A and B component. The brightest part (B) is usually referred to as 3C\,225 in literature. We use 3C\,225 to refer to 3C\,225B, unless explicitly stated otherwise.} is an FR II-type  double radio galaxy.
Previously this source sparked interest due to the lack of a bright optical counterpart, despite the compact nature of the radio emission \citep{compact_3C225}.
It has a redshift of 0.580 \citep{1985PASP...97..932S}.
The flux density of this source is a factor of $\sim$3 lower than the other sources, (see \autoref{tab:sources}), which makes this source a good proving ground for ULF imaging of relatively faint objects with the ILT.

Our observations, which reach a resolution of 1.1 arcseconds and a RMS noise of 58~\mjybeam, show that the low-frequency structure roughly resembles that seen at 5 GHz \citep{giovannini_1988}, with two lobes separated by $\sim 5 ''$.
The fainter component is about half as bright as the bright lobe and has a slightly extended feature, which is absent from the brighter lobe.
The usual explanation for the difference in brightness is the inclination of the jet with our line of sight, which causes significant Doppler beaming of the bright jet in our direction \citep[e.g.,][]{1989ApJ...336..606B}.
 In turn, the counter-jet would be beamed with a similar Doppler boosting factor $\beta$, causing it to be fainter by a factor \begin{equation}
F=\left[\frac{1+\beta\cos(\theta)}{1-\beta\cos(\theta)}\right]^{2-\alpha},\label{eq: betacos}
\end{equation} where $\theta$ is the inclination with respect to the line of sight and $\alpha$  the intrinsic radio spectral index, which we assume to be $\sim -0.83$.
Using \autoref{eq: betacos}, we find an estimate of $\beta\cos(\theta) \approx 0.124$.
Because there is no published estimate of either $\beta$ nor $\theta$, we are unable to provide any further constraints on either parameter.

\subsection{3C\,273}
\label{sec:3C273}
3C\,273 is the closest bright quasar \citep{1963Natur.197.1037H,1963Natur.197.1040S}, with a redshift of z=0.158 \citep{1992ApJS...83...29S}.
3C\,273 is radio-loud, variable, and relatively bright, which makes this object one of the most intensively studied quasars \citep[][]{1998A&ARv...9....1C}.
The source consists of two main parts: the core and an extended jet.
The core shows strong brightness changes on timescales of $\sim 10$ yr in the UV, radio, and optical regimes \citep{1992ApJ...396..469H}.
Due to its large physical size, the jet does not show this time-dependent brightness change.
The radio spectrum of the jet is relatively steep ($\sim -0.8$), and the jet is brighter than the core at frequencies below 300\,MHz \citep{2017A&A...601A..35P}.

\begin{figure*}
    \centering
    \includegraphics[width=\textwidth]{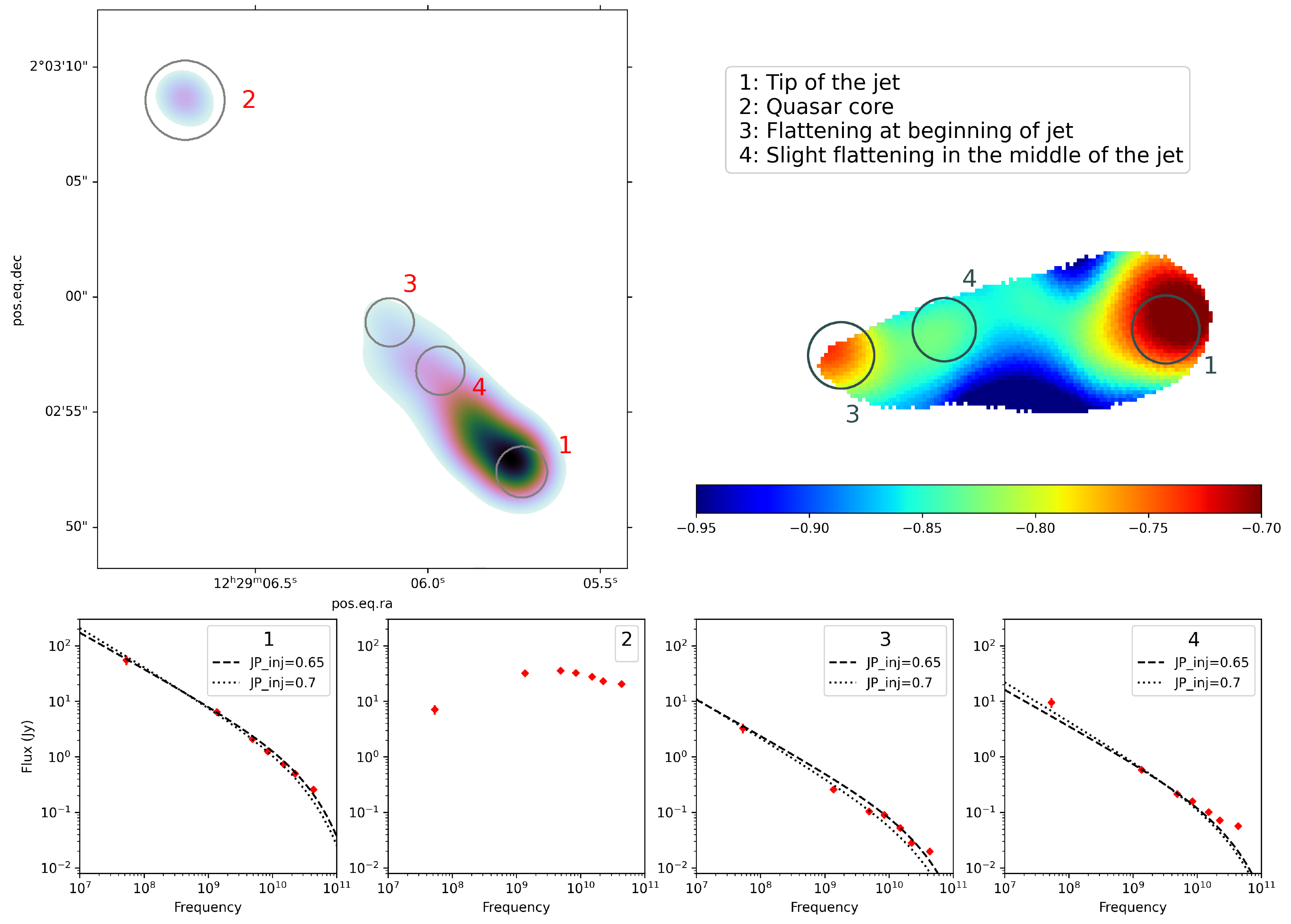}
    \caption{Spectra of four different locations of 3C\,273 using data from \cite{2017A&A...601A..35P}. The locations are indicated by circles in the 30--80~MHz images in the top left panel and the spectral index map in the top right panel. Overplotted are JP spectra \citep{jpmodel}, shifted to match the spectra. This means that it is not possible to extract an estimate of the age or magnetic field from this data alone. The VLA images and LOFAR LBA image are all smoothed to a similar circular beam before the flux densities in the indicated regions is determined. The flux densities tend to exceed the flux density given by the JP model.}
    \label{fig:fourspectra}
\end{figure*}

\begin{figure}
    \centering
    \includegraphics[width=1\linewidth]{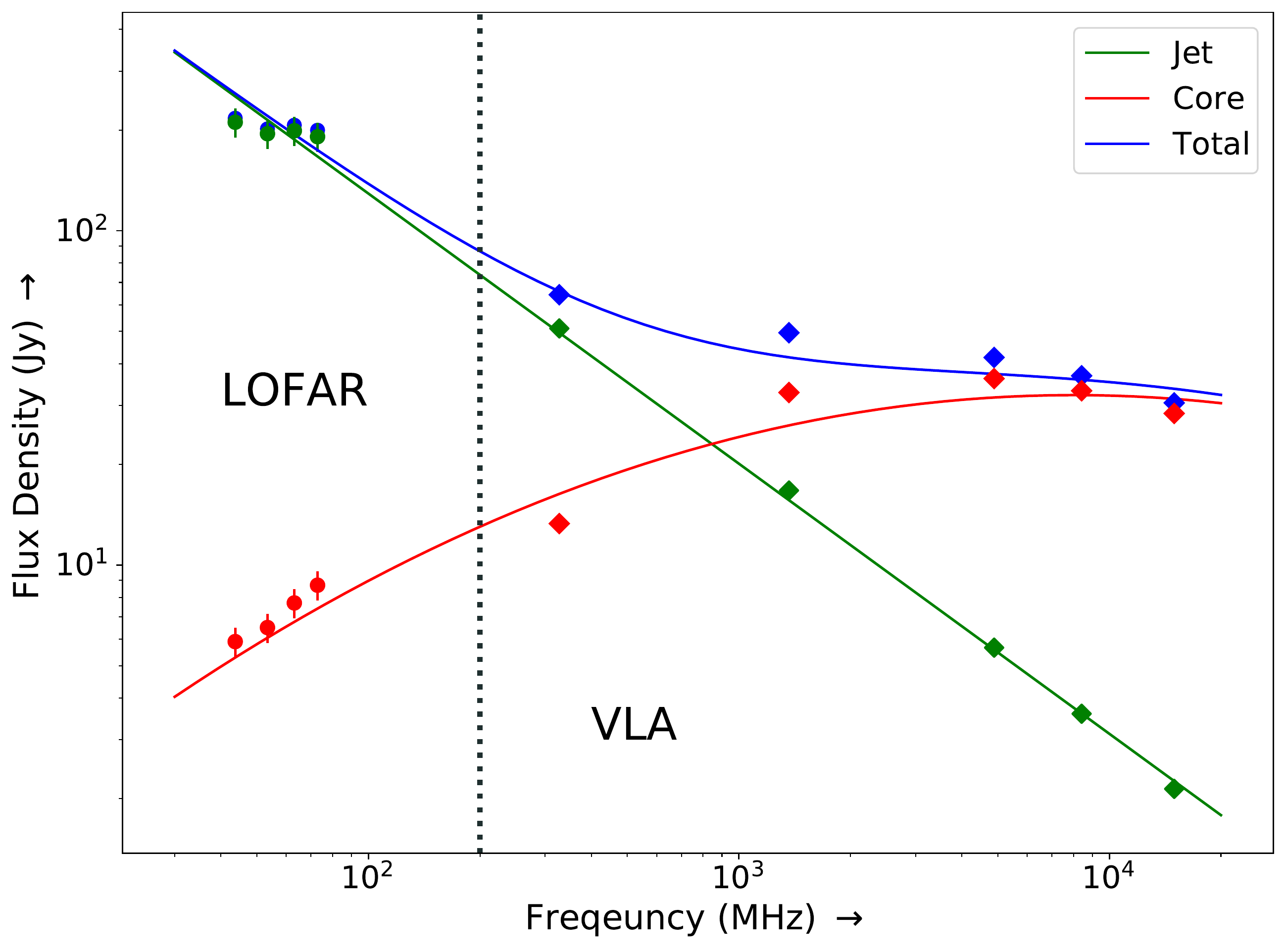}
    \caption{Flux densities determined from our LOFAR observations (circles) of 3C\,273, along with flux densities derived from the observations in \citet{2017A&A...601A..35P} (diamonds). The flux densities are split  into a core component (3C\,273 B, in red) and a jet component (3C\,273 A, in green), with the sum of the contribution of the jet and the core shown in  blue. The flux densities are measured directly from the resolved images of 3C\,273. The error bars for the VLA flux densities are smaller than the plotted points. The reduced chi-squared value of the fit for the jet is 1.43 and for the core 8.83.}
    \label{fig:ap_perleylofar}
\end{figure}

Our image of 3C\,273 is a significant improvement in spatial resolution over previous images at low frequencies \citep[e.g., compared to][which reached 4 arcsecond resolution at 328~MHz]{2017A&A...601A..35P}.
We obtain a resolution of 1.1~arcsecond with a RMS noise of 33~\mjybeam.
Our result reveals the substructure of the radio jet, and in turn allows us to constrain the physical parameters of this object.
The spectral index map in \autoref{fig:spixes} shows that the core has an inverted spectrum, while the jet has a typical spectral index of about -0.8, consistent with the findings of \cite{2017A&A...601A..35P}.
In addition, \autoref{fig:fourspectra} shows that the jet has two regions where the spectral index flattens (region 1 and 3 in \autoref{fig:fourspectra}).
This supports the hypothesis that the knotty structure is caused by internal shocks, where particles are (re-)accelerated, rather than beaming by the potential helical structure of the jet \citep[][]{mooney_3c273,harwood_3c273_papersplash}. 
There is another region of slight flattening of the jet spectrum, near C1 and C2 in \citet{hubble_nomenclature_3c273} and corresponding with region 4 in \autoref{fig:fourspectra}.
The substructure in the jet could also indicate previous cycles of AGN activity.

In \autoref{fig:ap_perleylofar} we show separately the total flux from the core (3C\,273\,B) and from the jet (3C\,273\,A).
Separate functions were fitted to the core and the jet, a first-order model to the jet, and a second-order polynomial model to the core.
At 150~MHz the jet has a spectral index of $-0.81$ and the core has a spectral index and curvature term of $0.53$ and $-0.15$, respectively.
The blue line, the total of the jet and core, has the functional form    
\begin{equation}
    F_\nu = 104~\text{Jy} \cdot \left(10^{0.53x-0.15x^2} + 10^{-0.81x}    \right)
    \label{eq: 3c273_model}
,\end{equation}
with $x=\log\left(\frac{\nu}{150\text{MHz}}\right)$.
Using the fit, we estimate a flux density of 104~Jy at 150~MHz.
The functional form of \autoref{eq: 3c273_model} is used as the reference model of 3C\,273 in \autoref{fig:comp_cats}.

Another peculiar property of this source is the lack of a counter-jet.
With our LBA observations, we do not find any significant flux in the region to the northeast of the core, where we expect the counter-jet to be.
The lack of a counter-jet is usually caused by Doppler boosting, where 3C\,273A (the bright radio jet) is significantly boosted in our direction, and the counter-jet is beamed away from us.
Assuming a spectral index of $\alpha \sim -0.8$ and using \autoref{eq: betacos}, a lower limit on the inclination can be estimated.
Taking the RMS noise of $0.033$~Jy $\text{beam}^{-1}$ on our LBA image and a peak flux at the tip of the jet of $40~ \text{Jy beam}^{-1}$ (which corresponds to the brightest value of the flux density in 3C\,273), this gives a lower limit on the ratio of the observed flux of the jet to the counter-jet of at least 242:1, assuming the counter-jet is fainter than the 5$\sigma$ noise level and under the assumption that the two jets have a similar emissivity.
Using \autoref{eq: betacos}, this gives us a lower limit of $\beta\cos(\theta) \geq 0.753$.
Assuming a jet velocity of $v=0.9c$, we obtain an estimate of the upper limit of the inclination of $\theta \approx \SI{33}{\degree}$, in agreement with \cite{conway_3c273}.

\subsection{3C\,295}
3C\,295 is a luminous radio galaxy with redshift z=0.46 \citep{2000AJ....119..451L} and an FR II morphology, with two bright radio lobes.
This object is located in a cluster \citep{1981ApJ...251..485M} with a weak cool core \citep{2001MNRAS.324..842A}, indicating that the cluster is not undergoing a major merger.

X-ray emission has been detected from the 3C\,295 cluster, its core \citep{2000ApJ...530L..81H}, and its radio lobes \citep{lobes_295_xray}.
VLA radio observations \citep{1991AJ....101.1623P} show a large change in rotation measure ($-7000 \text{ to } 9000\text{ rad m}^{-2}$) across the southeastern lobe, which indicates a substantial change in the direction of the magnetic field. 
Further observations by \citet{1992A&A...262..417T} show that the NW and SE lobes significantly differ in both structure and spectral shape.
The radio spectrum of this source has  a steep spectrum above 100 MHz, with a turnover at about 60 MHz.
However, explaining this by synchrotron self-absorption is challenging as this would require a strong magnetic field ($\sim$1~G), and the sharpness of the spectral turnover excludes a low-energy cutoff.
This makes free-free absorption a plausible cause of the turnover \citep{1992A&A...262..417T}.
The turnover makes this source a challenging target for ultra-low-frequency observations as the source becomes relatively faint below 30 MHz.

LOFAR HBA images have yielded high-resolution images, which we  used as our starting model for self-calibration \footnote{Made by F. Sweijen, see the skymodels in \url{https://github.com/lofar-astron/prefactor/}}. 
The LBA image, which has a resolution of 0.65~arcsecond with a RMS noise of 17~\mjybeam, is presented in \autoref{fig:full_plots}, and shows a similar structure to the HBA image.

\subsection{3C\,298}
3C\,298 is a quasar with two bright radio lobes, at a redshift of z=1.436 \citep{1993ApJ...413..453N}.
The host galaxy is currently in an intermediate stage merger, and is an interesting target for studying AGN feedback on galactic scales \citep{2017ApJ...851..126V}.
In addition, there is thermal emission in a region surrounding the optical core, detected at 1~mm, suggesting that significant star formation is occurring in the circumnuclear region surrounding the quasar nucleus \citep{2018ApJ...866L...3B}.

Our ILT image, which has a resolution of 1.0~arcsecond with a RMS noise of 215~\mjybeam, shows two lobes separated by approximately $1.5 
\text{''}$, the western lobe being brighter than the eastern lobe.
The western lobe has a flux density of 54\,Jy and the eastern lobe has a flux density of about 23\,Jy at 55 MHz.
The shape of the source resembles that derived from VLA observations at 1465 MHz by \citet{akujor_1995}.

\subsection{3C\,380}
3C\,380 is a radio galaxy at a redshift of 0.692 \citep{redshift3C380}.
The object consists of a compact steep spectrum (CSS) core, which is unresolved with LOFAR LBA, and a more diffuse extended structure, which is resolved on arcsecond scales \citep[see, e.g.,][]{perleybutler2017}.
3C\,380 is a commonly used source for calibration of LOFAR LBA observations as this source has a steep spectrum down to 10\,MHz, although an accurate model  is still lacking.

Our image, which has a resolution of 1.2~arcsecond and a RMS noise of 38~\mjybeam, confirms that the CSS core is enveloped in extended emission.
The diffuse halo is $\sim 60$ kpc in size, matching the observations of \citet{3c380_vlamerlin}.
In addition, there appears to be a curved structure of unknown nature on the eastern side of the core, which is also visible at 16~GHz \citep{perleybutler2017}.

\begin{figure*}
    \centering
    \includegraphics[width = 0.95\linewidth]{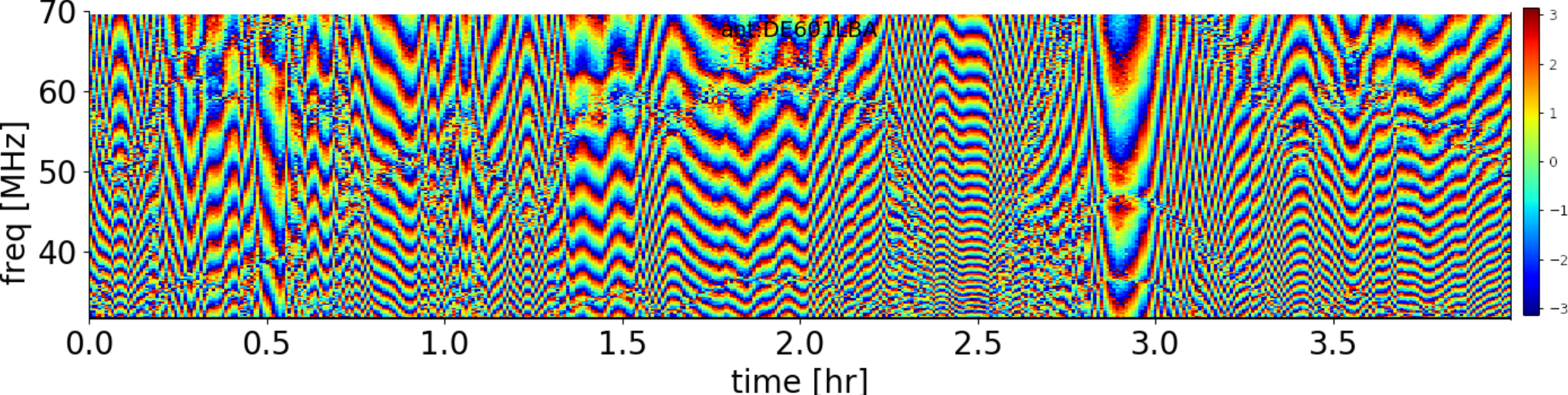}
    \caption{Overview of the XX phase corrections on station DE601LBA (Effelsberg, Germany), relative to the phased-up core, of a four-hour scan of 3C\,380. The scan is divided into time and frequency bins (in this case, 32~s wide bins in time space and 195~kHz in frequency space). In each bin, \texttt{DPPP} computes a single value for the phase correction for each polarization and for each antenna. The x-axis labels the time and the y-axis labels the frequency (in MHz) for each bin. The colormap represents the correction in radians, ranging from $-\pi$ to $\pi$. From this image, we can determine the dTEC value to be about $\sim\SI{0.5e16}{\per\meter\squared}$, referenced to the phased-up core \citep[see, e.g.,][]{2016ApJS..223....2V}. In the frequency direction the correction wraps around many times, depending on the severity of ionospheric conditions. The width of the bins is chosen to be as broad as possible (to increase the signal-to-noise ratio in each bin), while still ensuring that the correction does not change appreciably.}
    \label{fig:example_corrections}
\end{figure*}

\section{Future prospects}
\label{sec:discussion}
The results presented in this paper serve as a proof of concept for calibration of the ILT for frequencies down to 30~MHz. For example, the observations of 3C\,273, as presented in this paper, show that the ability of LOFAR LBA to match the resolutions obtained with the VLA at 2.4 GHz, which allows us to make spectral index maps with unprecedented high resolution at ULF. %The main limitation is the calibration of ionospheric effects. 
These spectral index maps reveal previously unknown old populations of relativistic electrons and spatially resolve these populations at arcsecond resolution.

High-resolution imaging below 80~MHz has three challenges.
The first of these is the variability of the ionosphere. This will make it more difficult to extend this work to fainter sources and to lower frequencies. 

To remove ionospheric effects we need to correct for rapidly varying phase effects on timescales of seconds and achieve a sufficiently high signal-to-noise ratio on these timescales to use self-calibration. Furthermore, the rapid variations of the phases as a function of frequency limit the bandwidth that can be used to derive a single solution.
In \autoref{fig:example_corrections} we show the phase corrections for a four-hour scan of 3C\,380.
Especially at lower frequencies, the solution interval in time and frequency is too large; at certain times (e.g., at about the 2.5~hour mark) there are fewer than six samples along the frequency axis per full rotation of the phase solutions.
However, decreasing the interval size will make solutions too noisy to be useful.
In addition, because the LOFAR station beam model is not fully accurate, amplitude corrections are often also needed, although on longer timescales of about an hour.
For this reason the signal-to-noise ratio achieved using the present ILT may be insufficient for the amplitude and phase calibration of fainter sources.
This is especially problematic for sources that have a low-frequency spectral turnover, which leads to even lower flux at lower frequencies.
A possible solution is to separate the clock and total electron content effects (clock--TEC separation) \citep{2019A&A...622A...5D}, and solve for each of these effects separately.
The frequency dependency of clock and TEC effects is known (clock $\sim\nu$, TEC $\sim\nu^{-1}$), which allows direct fitting of TEC and clock effects to the data.
Compared to the current approach, for which the data is divided into frequency bins, where the correction is assumed to be constant, clock--TEC separation allows the use of the full bandwidth to determine the corrections.
Clock--TEC separation may be useful in future ULF projects, although this would require the current separation algorithm to be adopted for usage with low signal-to-noise ratios.
The gain in image quality is limited for bright sources, such as the six sources discussed in this paper.
For fainter sources, however, clock--TEC separation is critical for obtaining useful results.

The second challenge is the lack of availability of good primary calibrators for ULF sources.
For primary calibrators, good high-resolution models are required for accurate extraction of the bandpass and polarization alignment, and  currently only 3C\,196 has a model that is sufficiently accurate at frequencies below 80~MHz.
Because the images that have resulted from this pilot project have unprecedently high resolution, they can be used for future Dutch LOFAR calibration reference models at frequencies below 80~MHz.
For 3C\,380, only a low-resolution model existed, consisting of a single point source.
Our results for 3C\,380 yield the first subarcsecond calibration reference model at low frequencies.
However, because the signal on the long baselines was insufficient to achieve long baseline calibration of sufficient quality, this model is not good enough for bandpass or polarization alignment calibration for the full International LOFAR Telescope, although it is a significant improvement for the Dutch LOFAR Telescope.
Our results for 3C\,295 are also an improvement over the previous two-component model for low frequencies and adequate for use on the Dutch and nearest international LOFAR stations.

The third challenge is the influence of solar activity on ionospheric conditions.
High solar activity is linked to severe ionospheric conditions, which make the calibration of large baselines especially difficult.
The data in this work were observed under good ionospheric circumstances.
A systematic investigation is necessary in order to understand the effect of elevation, sunset, and the solar cycle on the quality of the recorded data.
It is unclear what the severity of the impact of the solar cycle will be on the ability to correct for differential ionospheric effects.

\section{Conclusions}
\label{sec:conclusions}
We have shown that LBA data from the full International LOFAR Telescope can be used to generate subarcsecond images down to 30~MHz on six relatively bright sources.
In order to calibrate our data, we used bandpass responses and polarization alignments extracted from 3C\,196, after which we used several iterations of phase-only self-calibration, followed by several iterations of diagonal (complex gain) self-calibration.
Our results show that LOFAR LBA imaging with the international baselines is feasible, but that observations are currently limited to bright and compact objects.
The quality of the results seems to be sufficient to map the extended structure of the sources with a dynamic range of at least 100.\\

The resulting images are presented individually, along with spectral index maps of 3C\,196 and 3C\,273.
These results showcase the first subarcsecond results down to 30~MHz, and allow us to match the resolution of current gigahertz frequency observations.
This, in turn, allows us to extract spatially resolved spectral information.
A spectral index map of 3C\,196 shows that the east--west lobes contain older plasma than the north--south lobes, and in addition, that the source is embedded in a region of steep-spectrum emission.
The spectral index map of 3C\,273 shows that the knots visible in X-band and Hubble Space Telescope data have relatively flat spectra, which supports the hypothesis that these are probably caused by internal shocks.
Our observations of 3C\,380 have resolved the source for the first time below 80 MHz, revealing the halo structure surrounding the core.

It is important to explore ways of improving the sensitivity of the ILT at ULF.
One way of doing this could be by adding judiciously distributed collecting area (e.g., additional LBA stations) to improve the uniformity of the baseline coverage and facilitate the creation of additional large ``superterps'' \citep{lofar_paper}.
Further analysis of the optimum configuration of additional antenna stations and their effect on ULF calibration is desirable.
Because of the wealth of physical diagnostics accessible in the ULF and the demonstrated technical feasibility of ULF imaging with the ILT, a more sensitive ULF ILT would provide a unique capability for astronomy.

\begin{acknowledgements}
    We kindly thank the anonymous referee for the valuable and instructive comments.
    CG and RJvW acknowledge support from the ERC Starting Grant ClusterWeb 804208.
    LKM is grateful for support from the UKRI Future Leaders Fellowship (grant MR/T042842/1).
    AB acknowledges support from the VIDI research programme with project number 639.042.729, which is financed by the Netherlands Organisation for Scientific Research (NWO).
    JM acknowledges financial support from the State Agency for Research of the Spanish MCIU through the ``Center of Excellence Severo Ochoa'' award to the Instituto de Astrof\'isica de Andaluc\'ia (SEV-2017-0709) and from the grant RTI2018-096228-B-C31 (MICIU/FEDER, EU)
    LOFAR is the LOw Frequency ARray designed and constructed by ASTRON. It has observing, data processing, and data storage facilities in several countries, which are owned by various parties (each with their own funding sources), and are collectively operated by the ILT foundation under a joint scientific policy. The ILT resources have benefitted from the following recent major funding sources: CNRS-INSU, Observatoire de Paris and Universit\'{e} d'Orl\'{e}ans, France; BMBF, MIWF-NRW, MPG, Germany; Science Foundation Ireland (SFI), Department of Business, Enterprise and Innovation (DBEI), Ireland; NWO, The Netherlands; The Science and Technology Facilities Council, UK; Ministry of Science and Higher Education, Poland; Istituto Nazionale di Astrofisica (INAF), Italy.
    This research made use of Astropy,\footnote{http://www.astropy.org} a community-developed core Python package for Astronomy \citep{astropy:2013, astropy:2018}.
\end{acknowledgements}

\bibliographystyle{aa}
\bibliography{refs}

\end{document}